\documentclass[12pt]{article}

\InputIfFileExists{Debug}{}{}

\usepackage[utf8]{inputenc}
\usepackage{amsmath}
\usepackage{amsthm}
\usepackage{amsfonts}          % if you want the fonts
\usepackage{amssymb}           % if you want extra symbols
\usepackage[mathscr]{eucal}
\usepackage{graphicx}
\usepackage{color}
\usepackage{ulem}
\usepackage{hypbmsec}
\usepackage{fancyvrb}
\usepackage{sepnum}
\usepackage{xspace}

\usepackage{rotating}
\usepackage[isu,bf]{caption}
\usepackage{algorithm}
\usepackage{algorithmic}

\font\csc=cmcsc10

\newlength{\xtrawidth}
\newlength{\xtraheight}
\usepackage[all]{xy}           % Commutative diagrams

\ifx\NoBackreferences\EMPTYMACRO
  \usepackage[backref,linktocpage,bookmarks]{hyperref}
\else
  \usepackage[linktocpage,bookmarks]{hyperref}
\fi

\ifx\DEBUG\EMPTYMACRO
\else
\fi

\def\clap#1{\hbox to 0pt{\hss#1\hss}}
\def\mathllap{\mathpalette\mathllapinternal}

\def\mathllapinternal#1#2{%
\llap{$\mathsurround=0pt#1{#2}$}}

\makeatletter
  \def\adots{\mathinner{\mkern2mu\raise\p@\hbox{.}
      \mkern2mu\raise4\p@\hbox{.}\mkern1mu
      \raise7\p@\vbox{\kern7\p@\hbox{.}}\mkern1mu}}
\makeatother

\newcommand{\comma}[1]{\ensuremath{\sepnum{{.}}{{,}}{}{#1}}}

\newcommand{\eqdef}{%
  \mathrel{\lower.1mm
    \hbox{$\stackrel{\lower.424ex\hbox{\scriptsize def}}{=}$}}
}

\newcommand{\C}{\ensuremath{{\mathbb{C}}}}

\newcommand{\Z}{\mathbb{Z}}
\newcommand{\CP}{{\ensuremath{\mathop{\null {\mathbb{P}}}\nolimits}}}

\newcommand{\rt}{\ensuremath{\tilde{r}}}

\DeclareMathOperator{\Span}{span}
\DeclareMathOperator{\Pic}{Pic}

\DeclareMathOperator{\Tr}{Tr}

\DeclareMathOperator{\Sym}{Sym}
\DeclareMathOperator{\Alt}{Alt}

\DeclareMathOperator{\img}{img}
\DeclareMathOperator{\coker}{coker}

\DeclareMathOperator{\Hom}{Hom}

\DeclareMathOperator{\diag}{diag}

\DeclareMathOperator{\Ind}{Ind}
\DeclareMathOperator{\Res}{Res}
\DeclareMathOperator{\AltInd}{AltInd}
\DeclareMathOperator{\SymInd}{SymInd}
\DeclareMathOperator{\GrInd}{GrInd}

\DeclareMathOperator{\Stab}{Stab}

\newcommand{\Xt}{{\ensuremath{\widetilde{X}}}}

\newcommand{\Lsheaf}{\ensuremath{\mathscr{L}}}
\newcommand{\Osheaf}{\ensuremath{\mathscr{O}}}

\newcommand{\Vsheaf}{\ensuremath{\mathscr{V}}}

\newcommand{\tmod}{~\mathrm{mod}~}

{\end{list}}

{\everymath{\displaystyle\everymath{}}\array}%
{\endarray}

\newtheorem{theorem}{Theorem}
\newtheorem{lemma}{Lemma}
\newtheorem{example}{Example}

\newtheorem{definition}{Definition}

\begin{document}
%%%%%%%%%%%%%%%%%%%%%[ Title Page ]%%%%%%%%%%%%%%%%%%%%%%%%%%
\begin{titlepage}
  \vspace*{-2cm}
  \hfill
  \parbox[c]{5cm}{
    \begin{flushright}
%       arXiv:yymm.nnnn [hep-th]
%       \\
      DIAS-STP 10-03
    \end{flushright}
  }
  \vspace*{\stretch1}
  \begin{center}
    \Huge
    On Free Quotients of\\
    Complete Intersection
    Calabi-Yau Manifolds
  \end{center}
  \vspace*{\stretch1}
  \begin{center}
    \begin{minipage}{\textwidth}
      \begin{center}
        {\csc Volker Braun}
        \\[3ex]
        \textit{Dublin Institute for Advanced Studies}\\
        \textit{10 Burlington Road}\\
        \textit{Dublin 4, Ireland}\\[3ex]
        Email: \texttt{vbraun@stp.dias.ie}
      \end{center}
    \end{minipage}
  \end{center}
  \vspace*{\stretch1}
  \begin{abstract}
    In order to find novel examples of non-simply connected Calabi-Yau
    threefolds, free quotients of complete intersections in products of
    projective spaces are classified by means of a computer
    search. More precisely, all automorphisms of the product of
    projective spaces that descend to a free action on the Calabi-Yau
    manifold are identified.
  \end{abstract}
  \vspace*{\stretch1}
\end{titlepage}
\tableofcontents

\section{Introduction}
\label{sec:Intro}

Almost all Calabi-Yau manifolds that we know about are simply
connected. For example, the largest known class of Calabi-Yau
threefolds was classified in~\cite{KreuzerSkarkeReflexive,
  Kreuzer:2002uu} and consists of 3-d hypersurfaces in 4-d toric
varieties. The ambient toric varieties correspond to (usually
numerous) subdivisions of the normal fans of \comma{473800776}
reflexive 4-d polyhedra. Only $16$ of those lead to Calabi-Yau
hypersurfaces with non-trivial fundamental
group~\cite{Batyrev:2005jc}, which moreover ends up being either
$\pi_1(X)=\Z_2$, $\Z_3$, or $\Z_5$.

Are non-simply connected Calabi-Yau manifolds genuinely rare or is
this simply a case of ``searching under the lamppost''?  Note that, to
each non-simply connected manifold $X$, there is associated a unique
simply connected manifold, its universal cover $\Xt$, with a free
$\pi_1(X)$ action. Moreover, by modding out this free action we can
recover the original manifold. This suggests that one should search
for free actions on already known Calabi-Yau manifolds in order to
find new ones with non-vanishing fundamental group. This approach has
been successful for a long time~\cite{Yau1, Yau2, Greene:1986jb,
  Greene:1986bm}, and produced quite a number of manifolds of
phenomenological interest for heterotic string compactifications. 

A very convenient subset of (simply-connected) Calabi-Yau manifolds
are the $7890$ complete intersections in products of projective spaces
(CICY). Not only are they small enough in number to be easily handled
with a modern computer, but their ambient spaces also come with a
rather evident automorphism group. They have been a source for free
group actions for a long time~\cite{Candelas:1987kf, Candelas:1987du,
  Gross2000, szendroi-2001}. In a painstaking manual search~\cite{SHN}
most of the free group actions were actually found. However, some
remained hidden including a very curious three-generation
manifold~\cite{Braun:2009qy} with minimal Hodge numbers $h^{11}(X)=1$,
$h^{21}(X)=4$. Another application of the free CICY quotients is that,
in contrast to the simply-connected CICYs, they contain examples of
ample rigid divisors that are useful for moduli
stabilization~\cite{Bobkov:2010rf}. In the remainder of this paper, we
will perform an exhaustive search through the automorphisms of
products of projective spaces and classify all that restrict to a
\emph{free} action on the complete intersection Calabi-Yau
threefolds. A similar search can be performed for more general
complete intersections in toric varieties, but we leave this for
future work.

Before delving into the classification, we would like to apologize to
the reader for the horrendous technicalities that lie ahead. It is
strongly recommended to start with the results in
\autoref{tab:CICYgroups} on page~\pageref{tab:CICYgroups} and their
discussion in \autoref{sec:CYgroups}. The list of all free group
actions is included in the source code of this paper which can be
obtained from the arXiv server, see \autoref{sec:data} for more
details.

\section{The Classification}
\label{sec:classification}

\subsection{CICY Group Actions}
\label{sec:cicy}

The goal of this paper is to classify group actions on Calabi-Yau
threefolds that are complete intersections in products of projective
spaces (CICY). Moreover, we will only consider group actions that come
from group actions on the ambient space $\prod_i \CP^{d_i-1}$. That is,
we only consider group actions that are combinations of
\begin{enumerate}
\item Projective-linear action on the individual factors $\CP^{d_i-1}$,
  and
\item Permutations\footnote{Called \emph{external}
    in~\cite{Candelas:1987du}, but we will not use this notation in
    the following.} of the factor $\CP^{d_i-1}$.
\end{enumerate}
In other words, we only allow group actions that are represented by
linear transformations on the combined homogeneous coordinates. These
are also the group actions of physical interest for the construction
of (equivariant) monad bundles, see~\cite{Anderson:2008uw,
  Braun:2009mb, He:2009wi, Anderson:2009mh}. In general, there are
also non-linear group actions. However, in special cases we classify
actually all possible group actions. For example, when the Calabi-Yau
manifold in question is given by its Kodaira embedding\footnote{That
  is, there is a (invariant but not necessarily equivariant) line
  bundle $\Lsheaf$ on $X$ such that the $d=h^0(X,\Lsheaf)$ global
  sections $s_\alpha$ do not vanish simultaneously and separate points
  and tangent directions. That is, $x\mapsto [s_0(x):\cdots:s_d(x)]$
  defines an embedding into $\CP^{d-1}$.}  $X\subset \CP^{d-1}$, then
all actions are linear. In particular, any group action on the Quintic
in $\CP^4$ is of the type we are considering.

Recall the standard notation for the degrees of the transverse
polynomials defining a CICY manifold. This is just a matrix $(c_{ij})$
such that the $j$-th polynomial is of homogeneous degree $c_{ij}$ in
the homogeneous coordinates of the $i$-th projective space. For the
group action to descend to the complete intersection the individual
polynomials need not be preserved, only their common zero set must
be. In particular, if multiple polynomials of the same degree occur
then they might be transformed into non-trivial linear combinations.

This is why we will use a slightly different notation where the
degrees (and, hence, the diffeomorphism type) of the CICY is defined
by a configuration matrix with pairwise different columns
\begin{equation}
  \renewcommand{\arraystretch}{1.7}
  \begin{tabular}{c|c|c|c|c|} 
    \multicolumn{1}{c}{} & 
    \multicolumn{1}{c}{{$\vec{p}_1$}} &
    \multicolumn{1}{c}{{$\vec{p}_2$}} &
    \multicolumn{1}{c}{{$\cdots$}} &
    \multicolumn{1}{c}{{$\vec{p}_m$}} \\
    \cline{2-5} 
    {$\CP_1 \eqdef \CP^{d_1-1}$} & 
    $c_{11}$ & $c_{12}$ & $\cdots$ & $c_{1m}$ \\ \cline{2-5}
    {$\CP_2 \eqdef \CP^{d_2-1}$} & 
    $c_{21}$ & $c_{22}$ & $\cdots$ & $c_{2m}$ \\ \cline{2-5}
    {$\vdots$}
    & $\vdots$ & $\vdots$ & $\ddots$ & $\vdots$ \\ \cline{2-5}
    {$\CP_n \eqdef \CP^{d_n-1}$} & 
    $c_{n1}$ & $c_{n2}$ & $\cdots$ & $c_{nm}$ \\ \cline{2-5}
  \end{tabular}
  ~,
\end{equation}
meaning that 
\begin{itemize}
\item The ambient space is $\prod_{i=1}^n \CP_i$
\item The CICY is cut out by $m$ vectors of equations $\vec{p}_j$ each
  having $\delta_j\in \Z_>$ components.
\item Each component of the equation vector $\vec{p}_j$ is a
  homogeneous polynomial of degree $c_{ij} \in \Z_\geq$ in the
  $d_i$ homogeneous coordinates of the $i$-th factor $\CP_i$.
\end{itemize}
Obviously $n-3=\sum \delta_j$ for threefolds. Moreover, the vanishing
of the first Chern class is equivalent to
\begin{equation}
  d_i = \sum_{j=1}^m c_{ij} \delta_j
  \qquad \forall i=1,\dots,n
  .
\end{equation}
However, the group and index theory we will use is independent of the
dimension and Chern class and could be applied to more general
complete intersections.

To formalize this notion of group action, let us define
\newcommand{\Prow}{\ensuremath{P_\text{row}}}
\newcommand{\Pcol}{\ensuremath{P_\text{col}}}
\begin{definition}[CICY groups]
  \label{def:CICYgroup}
  A CICY group is a quadruple $(C,G,\pi_r, \pi_c)$
  where
  \begin{itemize}
  \item $C = (d_i,c_{ij},\delta_j)_{i=1..n,~j=1..m}$ is the
    configuration matrix of a CICY,
  \item $G$ is a group, 
  \item $\pi_r:G\to \Prow$ is a permutation action on the $n$ rows,
    and
  \item $\pi_c:G\to \Pcol$ is a permutation action on the  
  $r$ columns
  \end{itemize}
  such that the configuration matrix is invariant under the
  permutations. That is,
  \begin{equation}
    c_{i,j} = c_{\pi_r(g)(i), ~ \pi_c(g)(j)}
    \quad
    \forall g\in G
    ,
  \end{equation}
  the number $d_i$ of homogeneous coordinates of $\CP_i=\CP^{d_i-1}$
  is constant on orbits of $\Prow$, and the number of components
  $\delta_j$ of $p_j$ is constant on orbits of $\Pcol$.
\end{definition}
\begin{lemma}
  $\pi_c$ is uniquely determined by $(C,G,\pi_r)$ if it
  exists.   
\end{lemma}

Now, a \emph{representation} of a CICY group is the collection of
matrices, one for each group element and each projective space, acting
on the homogeneous coordinates. One must ensure that permutations
interchange the different projective space and equation vectors. Note
that this is the same structure for the rows and columns. Therefore,
let us define
\begin{definition}[$\pi$-representation]
  \label{def:pirep}
  A (linear) $\pi$-representation is a quadruple
  $(G,\pi,\vec{d},\gamma)$ where
  \begin{itemize}
  \item $\pi: G\to P$ is a permutation action of $G$ on
    $\{1,\dots,n\}$.
  \item $\gamma_i : G\to GL(d_i, \C)$ is a map satisfying
    \begin{equation}
      \gamma_{\pi(h)(i)}(g) \gamma_i(h) = \gamma_i (gh)
      \qquad 
      \forall g,h\in G
      ,~
      i \in \{1,\dots,n\}
      .
    \end{equation}
  \end{itemize}
\end{definition}
In other words, the $d_i\times d_i$ matrices $\gamma_i(g)$ can be
assembled into an ordinary group representation by block matrices
$\gamma(g)$ of the form
\begin{equation}
  \label{eq:gammadef}
  \gamma(g) = 
  \mathbf{P}\big(\pi(g), \vec{d}\big)
  ~
  \diag\big( \gamma_1(g),~ \dots,~ \gamma_n(g)\big)
\end{equation}
where $\mathbf{P}(\pi(g), \vec{d})$ is the permutation matrix
corresponding to the permutation $\pi(g)$ acting on $\{1,\dots,n\}$
but with entries being rectangular matrices $\mathbf{0}_{d_i\times
  d_j}$ and (square) identity matrices $\mathbf{1}_{d_i\times d_j}$
instead of $0$ and $1$.

A projective $\pi$-representation $(G,\pi,\gamma)$ is one where
$\gamma_i: G\to PGL(d_i)$. This is the case of interest to us, since
homogeneous coordinates as well as the zero sets of polynomials do not
depend on overall $\C^\times$ factors. 

Let us formalize the data required to define a group action on a CICY
manifold;
\begin{definition}[CICY group action]
  \label{def:CICYgroupaction}
  A CICY group action is a tuple $(C,G,\pi_r,\gamma,\pi_c,\rho)$ such
  that
  \begin{itemize}
  \item $C = (d_i,c_{ij},\delta_j)_{i=1..n,~j=1..m}$ is the
    configuration matrix of a CICY,
  \item $(C,G,\pi_r,\pi_c)$ is a CICY group, and 
  \item $(G,\pi_r,\vec{d},\gamma)$ and $(G,\pi_c,\vec{\delta},\rho)$
    are $\pi$-representations.
  \end{itemize}
\end{definition}
A CICY group action defines an action on the combined homogeneous
coordinates 
\begin{equation}
  \vec{z} 
  ~\eqdef
  \big[
  z_{1,1}:\cdots:z_{1,d_1}
  \big|
  z_{2,1}:\cdots:z_{2,d_2}
  \big| ~\cdots~ \big|
  z_{n,1}:\cdots:z_{n,d_n}
  \big]  
\end{equation}
of $\prod \CP_i$. This action induces a $\pi$-representation on
the combined polynomial equations
\begin{equation}
  \vec{p} 
  ~\eqdef
  \big( \vec{p}_1,~ \dots,~ \vec{p}_m \big) 
  =
  \big(
  p_{1,1},~ \dots,~ p_{1,\delta_1} ;~
  \dots ;~
  p_{r,1},~ \dots,~ p_{r,\delta_m}
  \big)
  .
\end{equation}
We say that the polynomials defining the CICY are \emph{invariant}
under the group action if this induced action on the equations equals
the representation $(G,\pi_c,\rho)$. In other words, the composition
\begin{equation}
  \rho^{-1}(g) ~ \vec{p}\Big( \gamma(g) \vec{z} \Big) 
  = 
  \vec{p}(\vec{z})
  \qquad 
  \forall \gamma \in G
\end{equation}
leaves the polynomials invariant. That is, the $(G,\pi_c,\rho)$ action
cancels out the non-trivial action on the polynomials. 
\begin{theorem}
  Fix a CICY group $(C,G,\pi_r, \pi_c)$, and a projective
  $\pi$-representation $(G,\pi_r,\vec{d},\gamma)$ acting on the
  homogeneous coordinates. Then the zero set $\{\vec{p}=0\}\subset
  \prod \CP_i$ is invariant if an only if there is a CICY action
  $(C,G,\pi_r,\gamma,\pi_c,\rho)$ leaving the polynomials
  invariant.
\end{theorem}
Finally, note that the invariant polynomials can be easily computed by
the usual Reynolds operator, that is, summing over orbits of the
group. 
\begin{example}[A CICY group action]
  Consider the CICY \#20, 
  \begin{equation}
    \left[
      \begin{tabular}{c|cccc}
        $\CP^1$ &   $0$ & $0$ & $1$ & $1$ \\
        $\CP^1$ &   $0$ & $0$ & $0$ & $2$ \\
        $\CP^1$ &   $0$ & $0$ & $0$ & $2$ \\ 
        $\CP^4$ &   $2$ & $2$ & $1$ & $0$ 
      \end{tabular}
    \right]
    \quad = \qquad
    \renewcommand{\arraystretch}{1.7}
    \begin{tabular}{c|c|c|c|} 
      \multicolumn{1}{c}{} & 
      \multicolumn{1}{c}{{$
          \left(\begin{smallmatrix}
            p_1 \\ p_2
          \end{smallmatrix}\right)$}} &
      \multicolumn{1}{c}{{$\big( p_3 \big)$}} &
      \multicolumn{1}{c}{{$\big( p_4 \big)$}} \\
      \cline{2-4} 
      {$\CP_1 \eqdef \CP^{1}$} & 
      $0$ & $1$ & $1$ 
      \\ \cline{2-4}
      {$\CP_2 \eqdef \CP^{1}$} & 
      $0$ & $0$ & $2$
      \\ \cline{2-4}
      {$\CP_3 \eqdef \CP^{1}$} & 
      $0$ & $0$ & $2$
      \\ \cline{2-4}
      {$\CP_4 \eqdef \CP^{4}$} & 
      $2$ & $1$ & $0$ 
      \\ \cline{2-4}
    \end{tabular}
    \quad
    = C
    .
  \end{equation}
  In particular, the numbers of homogeneous coordinates corresponding
  to each row are $d=(2,2,2,5)$, and the numbers of equations
  corresponding to each column are $\delta=(2,1,1)$.

  Now, let us consider the group $\Z_4=\{1,g,g^2,g^3\}$ generated by
  $g$. One possible CICY group for the configuration matrix $C$ is
  $(C,G,\pi_r,\pi_c)$ with the permutation actions\footnote{I am using
    cycle notation for the permutations.}
  \begin{equation}
    \pi_r(g) \eqdef (2,3)
    , \qquad
    \pi_c(g) \eqdef ()
    .
  \end{equation}
  An example of a CICY group action is
  $(C,\Z_4,\pi_r,\gamma,\pi_c,\rho)$ with the representations
  generated by
  \begin{equation}
    \begin{split}
      \gamma(g) \eqdef&\;
      \begin{pmatrix}
        i & 0 \\ 0 & -i
      \end{pmatrix}
      \oplus
      \begin{pmatrix}
        0 & 0 & 1 & 0 \\ 
        0 & 0 & 0 & -1 \\ 
        1 & 0 & 0 & 0 \\ 
        0 & 1 & 0 & 0
      \end{pmatrix}
      \oplus
      \begin{pmatrix}
        1 & 0 & 0 & 0 & 0 \\ 
        0 & -1 & 0 & 0 & 0 \\ 
        0 & 0 & i & 0 & 0 \\ 
        0 & 0 & 0 & i & 0 \\ 
        0 & 0 & 0 & 0 & -i
      \end{pmatrix}
      ,\\
      \rho(g) \eqdef&\;
      \begin{pmatrix}
        1 & 0 \\ 
        0 & -1 
      \end{pmatrix}
      \oplus
      \begin{pmatrix}
        1    
      \end{pmatrix}
      \oplus
      \begin{pmatrix}
        i    
      \end{pmatrix}
      .
    \end{split}
  \end{equation}
  A basis for the invariant polynomial vectors is
  \begin{equation}
    \begin{array}{@{(}c@{,~}c@{,~}c@{,~}c@{~)}l@{}r@{\quad(}c@{,~}c@{,~}c@{,~}c@{~),}}
      0 & 0 & 0 & z_2 z_3 z_4 z_5 z_6 & , &&
      0 & 0 & 0 & z_2 z_3^2 z_6^2-z_2 z_4^2 z_5^2 \\
      0 & 0 & 0 & z_1 z_4^2 z_6^2 & , &&
      0 & 0 & 0 & z_1 z_3^2 z_6^2+z_1 z_4^2 z_5^2 \\
      0 & 0 & 0 & z_1 z_3^2 z_5^2 & , &&
      0 & 0 & z_2 z_{10} & 0 \\
      0 & 0 & z_2 z_9 & 0 & , &&
      0 & 0 & z_1 z_{11} & 0 \\
      0 & z_{10}^2 & 0 & 0 & , &&
      0 & z_9 z_{10} & 0 & 0 \\
      0 & z_9^2 & 0 & 0 & , &&
      0 & z_7 z_8 & 0 & 0 \\
      z_{11}^2 & 0 & 0 & 0 & , &&
      z_{10} z_{11} & 0 & 0 & 0 \\
      z_9 z_{11} & 0 & 0 & 0 & , &&
      z_8^2 & 0 & 0 & 0 \\
      z_7^2 & 0 & 0 & 0 & .
    \end{array}
  \end{equation}
  One can show that a sufficiently generic linear combination cuts out
  a smooth fixed-point free CICY threefold.
\end{example}

\subsection{Classification Algorithm}
\label{sec:algo}

Using index theory one can show~\cite{Candelas:1987kf,
  Candelas:1987du} that any free group action on one of the $7890$
CICYs has $|G|\leq 64$. Hence, there are only a finite number of
possible CICY groups. Moreover, there is only a finite number of
distinct group representations for fixed dimension. Therefore, there
is only a finite number of free CICY group actions, and we
can, in principle, enumerate all of them:
\pagebreak[2]
\begin{algorithmic}[1]
\STATE \texttt{FreeActions} $= \{ \}$
\FORALL{CICY configuration matrices $C$} 
  \FORALL{CICY groups $(C,G,\pi_r,\pi_c)$ such that $|G|\leq 64$}
    \FORALL{$\pi$-representations $(G,\pi_r,\vec{d},\gamma)$ and
            $(G,\pi_c,\vec{\delta},\rho)$}
      \STATE $\vec{p} =$ random linear combination 
             of $(C,G,\pi_r,\gamma,\pi_c,\rho)$-invariant \rlap{polynomials}
% note the \rlap, algorithmic has some spacing issues
      \STATE $X = \{ \vec{p} = 0 \} \subset \prod \CP_i$
      \IF{$X$ is fixed-point free \textbf{and} $X$ is smooth}
        \STATE Add $(C,G,\pi_r,\gamma,\pi_c,\rho)$ to \texttt{FreeActions}
      \ENDIF
    \ENDFOR
  \ENDFOR
\ENDFOR
\RETURN \texttt{FreeActions}
\end{algorithmic}
\pagebreak[2]
Although finite, working through this algorithm is far out of reach of
present-day capabilities. Enumerating all $(G,\pi_r,\vec{d},\gamma)$
and all $(G,\pi_c,\vec{\delta},\rho)$ representations is feasible, but
their Cartesian product often exceeds $10^{10}$ pairs. Moreover,
checking for fixed points and, in particular, smoothness requires
Gr\"obner basis computations that can take from seconds to multiple
days on a modern desktop computer\footnote{All computations in this
  paper were done on a 2.66GHz Intel Core i7 processor and 12 GiB of
  RAM.} even using the algorithmic improvements outlined below.

The key to classifying the free actions is to compute the
character-valued indices of a sample of equivariant line
bundles. These must be of a certain ``free'' type, otherwise the group
action cannot be free on the CICY manifold. Moreover, these
character-valued indices can be computed without explicitly
constructing the representations or polynomials. In \autoref{sec:group}
we will introduce a generalization of Schur covers that is necessary
to compute characters of projective $\pi$-representations, and in
\autoref{sec:indices} we will show how to compute the indices using
character theory alone.

One still needs a few optimizations to classify all free CICY
quotients. These include
\begin{itemize}
\item Knowing the group $G$ lets us identify line bundles that must be
  equivariant. The ordinary (not character-valued) index must be
  divisible by $|G|$, yielding stronger restrictions than indices that
  only depend on the configuration matrix.
\item The $\pi$-representation $(G,\pi_r,\vec{d},\gamma)$ and
  $(G,\pi_c,\vec{\delta},\rho)$ can be decomposed into blocks
  corresponding to the $\img(\pi)$-orbits. The list of all ``big''
  representations is just the Cartesian product of all the
  representations corresponding to the individual $\img(\pi)$-orbits. 
\item In many CICYs there are a few line bundles whose
  character-valued index does not depend on all of the blocks of the
  $(G,\pi_r,\vec{d},\gamma)$ and
  $(G,\pi_c,\vec{\delta},\rho)$-representations. By testing these line
  bundles first, we can eliminate some choices for the contributing
  blocks without going through the whole Cartesian product. 
\item Smoothness and absence of fixed points can be checked much
  faster over finite fields. Choosing the wrong finite field or the
  wrong invariant polynomial may yield false negatives, but a positive
  answer is definite. By repeating the test with different finite
  fields and a different linear combination of invariant polynomials,
  we can make false negatives highly unlikely.
\item As we will show in detail in \autoref{sec:group}, one can
  enumerate the $(G,\pi_r,\vec{d},\gamma)$ and
  $(G,\pi_c,\vec{\delta},\rho)$-representations using characters. The
  explicit representation matrices are only required to check for
  fixed points and smoothness, but not to compute the character-valued
  indices.
\end{itemize}
\newcommand{\Gt}{\ensuremath{\widetilde{G}}}
\begin{algorithm}
\caption{Classifying the smooth free CICY quotients}
\label{alg:class}
\begin{algorithmic}[1]
\STATE \texttt{FreeActions} $ = \{ \}$
\FORALL{CICY configuration matrices $C$} 
  \FORALL{CICY groups $(C,G,\pi_r,\pi_c)$ such that\\\hfill $|G|$ divides all
    indices of $C$ \textbf{and} every subgroup of $G$ acts freely}
    \IF{topological index of some $G$-equivariant bundle is not
      divisible by $|G|$}
      \STATE continue with next CICY group
    \ENDIF

    \STATE Find generalized Schur cover $\Gt\to G$
    \STATE $\Gamma$ = all (linear)
    $\pi$-representations $(\Gt,\pi_r,\vec{d},\gamma)$
    \STATE $R$ = all (linear) 
    $\pi$-representations $(\Gt,\pi_c,\vec{\delta},\rho)$

    \FORALL{$\Lsheaf$ in a sample of invariant line bundles}
  
      \FORALL{$(\Gt,\pi_r,\vec{d},\gamma)\in \Gamma$ and
              $(\Gt,\pi_c,\vec{\delta},\rho)\in R$ 
              \\ \hfill that are not already ruled
              out by a previous line bundle $\Lsheaf$}

        \STATE Compute the character-valued index $\chi(\Lsheaf)$
        \IF{$\chi(\Lsheaf)$ is not of the free type}
          \STATE \COMMENT{$(C,\Gt,\pi_r,\gamma,\pi_c,\rho)$ cannot act freely}
          \STATE continue with next representation
        \ENDIF

        \STATE Compute the twist $\tau$, 
               a character of $\ker(\Gt\to G)$

        \IF[$\tau=1$ means $\Lsheaf$ is $G$-equivariant]
           {$\tau=1$ \textbf{and} $|G|\; \not\big|\; \dim\chi(\Lsheaf)$}
          \STATE continue with next representation
        \ENDIF

        \STATE Construct the
          explicit representation matrices for $\gamma$, $\rho$.

        \FOR{many finite fields $\mathbb{F}$}

          \STATE $\vec{p} =$ random $\mathbb{F}$-linear combination 
                 of $(C,\Gt,\pi_r,\gamma,\pi_c,\rho)$-invariants

          \STATE $X = \{ \vec{p} = 0 \} \subset \prod \mathbb{FP}_i$
          \IF{$X$ is fixed-point free \textbf{and} $X$ is smooth}
            \STATE Add $(C,G,\pi_r,\gamma,\pi_c,\rho)$ to \texttt{FreeActions}
          \ENDIF
        \ENDFOR
      \ENDFOR
    \ENDFOR
  \ENDFOR
\ENDFOR
\RETURN \texttt{FreeActions}
\end{algorithmic}
\end{algorithm}
Using these ideas, we present the improved Algorithm~\ref{alg:class}.
I implemented this classification algorithm using the GAP and
\textsc{Singular} computer algebra systems~\cite{GPS05, GAP4,
  GapSingular}. The whole program completed within a few months of run
time.

\section{Group Actions}
\label{sec:group}

\subsection{Projective Representations}
\label{sec:projective}

Recall that a (linear) representation of a group $G$ is a map
\begin{equation}
  r:G \to GL(n,\C)
  , \qquad
  r(g) r(h) = r(gh)
  \quad
  \forall g,h \in G
  .
\end{equation}
The matrices $r(g)$ clearly depend on the chosen basis, but
representations that merely differ by a coordinate transformation
should be regarded as the same. An obvious invariant of the
representation $r$ is its character
\begin{equation}
  \chi_r: G\to \C^\times
  ,~
  g\mapsto \Tr\Big( r(g) \Big)
  .
\end{equation}
Recall some well-known properties of the characters:
\begin{itemize}
\item $\chi_r(g)=\chi_r(h^{-1}gh)$ depends only on the conjugacy class
  of $g\in G$.
\item There is a one-to-one correspondence between irreducible
  representations and their characters.
\end{itemize}
Clearly, it is desirable to work with the characters instead of
(isomorphism) classes of representations. However, this requires that
all representations are linear, and not just projective.

Consider the following example of a projective representation,
\newcommand{\ZZxy}[2]{%
  \ensuremath{\left(
      \begin{smallmatrix}
        {#1} \\ {#2}
      \end{smallmatrix}
    \right)
  }}
\newcommand{\ZZpm}{\ZZxy{+}{-}}
\newcommand{\ZZmp}{\ZZxy{-}{+}}
\newcommand{\ZZpmt}{\ensuremath{\widetilde{\ZZpm}}}
\newcommand{\ZZmpt}{\ensuremath{\widetilde{\ZZmp}}}
\begin{example}[A projective representation]
  \label{example:projective}
  Let $G=\Z_2\times\Z_2 = \big\{\ZZxy{+}{+}, \ZZpm, \ZZmp,
  \ZZxy{-}{-}\big\}$ and
  \begin{equation}
    r\ZZpm=
    \begin{pmatrix}
      1 & 0 \\ 0 & -1
    \end{pmatrix}
    ,\quad
    r\ZZmp=
    \begin{pmatrix}
      0 & 1 \\ 1 & 0
    \end{pmatrix}
    .
  \end{equation}
  Thought of as $PGL(2)$-matrices, $r$ is a projective representation
  of $G$. However, the matrices $r\ZZpm, r\ZZmp\in GL(2)$ generate the
  group $D_8$, so $r$ is not a (linear) representation of
  $G$. Moreover, one cannot turn $r$ into a representation by
  multiplying $r\ZZpm$, $r\ZZmp$ by fixed overall phases.
\end{example}
As is obvious from the example, if one wants to work with linear
instead of projective representations one can lift them to linear
representations, but at the cost of having to enlarge the
group. Clearly, there is an epimorphism from the enlarged group $\Gt$
to the original group $G$ by making everything projective again. This
means that
\begin{equation}
  \label{eq:centralextension}
  1 
  \longrightarrow
  K
  \longrightarrow
  \Gt
  \longrightarrow
  G 
  \longrightarrow
  1
\end{equation}
is a \emph{central} extension, that is, the kernel $K$ is in the
center of $\Gt$. In other words, $K\subset \Gt$ are the commutators
that are non-trivial in $\Gt$ but become trivial when mapped into
$G$.

Thanks to Schur~\cite{35.0155.01, 38.0174.02} we know that, for any
finite group $G$, there is a finite covering group $\Gt$ such that
there is a one-to-many\footnote{Many because if $\psi:\Gt\to
  \C^\times$ is a one-dimensional representation then $\rt$ and
  $\psi \rt$ correspond to the same projective representation.}
correspondence between
\begin{itemize}
\item projective representations $r:G\to PGL(n)$ and
\item twisted representations, that is linear representations
  $\rt:\Gt\to GL(n)$ such that $\rt(k) \sim
  \mathbf{1}_{n\times n}$ for all $k \in K$.
\end{itemize}
Any such group is called a ``hinreichend erg\"anzte Gruppe''
(sufficient\footnote{Note that if $\Gt$ is a sufficient extension,
  then $\Gt\times H$ is sufficient as well. So there are infinitely
  many sufficient extensions.} extension) or of surjective type. If
$\Gt$ is of minimal size, then it is called a ``Darstellungsgruppe''
(representation group) or Schur cover. In general, a Schur cover is
not uniquely determined.

A twisted representation $\rt:\Gt\to GL(n)$ determines a
one-dimensional representation $\tau:K\to \C^\times$ via $\rt(k)
= \tau(k) \mathbf{1}_{n\times n}$. Multiplying $\rt$
with a one-dimensional representation of $\Gt$ also multiplies $\tau$,
so we should identify the orbits under this action. This leads to
\begin{definition}[Twist of a twisted representation]
  \label{def:twist}
  Consider a central extension eq.~\eqref{eq:centralextension} and let
  $\rt$ be a twisted representation. Then we say that\footnote{$\Hom$
    will always denote \emph{group} homomorphisms in this paper.}
  \begin{equation}
    \tau 
    = 
    \tfrac{1}{\dim \rt} \rt|_K
    =
    \tfrac{1}{\dim \rt} \Res^{\Gt}_K (\rt)
    \quad 
    \in 
    \Hom(K,\C^\times) \big/ \Res^{\Gt}_K\Hom(\Gt,\C^\times)
  \end{equation}
  is the ``twist'' of $\rt$. It is a one-dimensional
  representation of $K$ modulo the multiplicative action of the
  restrictions of one-dimensional representations of $\Gt$.
\end{definition}
In \autoref{sec:groupcohomology}, we will remark on the connection
between the twisted representations and the more standard approach
towards projective representations using group cohomology. However,
this is not necessary to understand the remainder of this paper.

Evidently, sums of representations with the same twist are again
twisted representations and correspond to a projective representation;
The sum of representations with different twists is not a twisted
representation. Finally, if $\tau=1$ is (equivalent to) the trivial
representation, then the corresponding projective representation is
actually linear.

\begin{example}[Continuation from Example~\ref{example:projective}]
  A Schur cover of $\Z_2\times\Z_2$ is $D_8$, leading to the central
  extension
  \begin{equation}
    1 
    \longrightarrow
    \left<
      \begin{pmatrix}
        -1 & 0 \\ 0 & -1
      \end{pmatrix}
    \right>
    \longrightarrow
    \left<
      \begin{pmatrix}
        1 & 0 \\ 0 & -1
      \end{pmatrix}
      ,~
      \begin{pmatrix}
        0 & 1 \\ 1 & 0
      \end{pmatrix}
    \right>
    \longrightarrow
    \Z_2 \times \Z_2
    \longrightarrow
    1
  \end{equation}
  The group $D_8$ has four 1-dimensional irreps (of twist $\tau=1$)
  and one 2-dimensional irrep of twist $\tau(-\mathbf{1}_{2\times
    2})=-1$.
\end{example}

\subsection{Induction and Restriction}
\label{sec:ind}

Using Schur covers and characters solves the problem of enumerating
all projective representations in an efficient manner. However, we
need to generalize it to representations in \emph{products} of
projective spaces where some group elements act by permutations.

For the reminder of this subsection, let us only consider
\emph{linear} $\pi$-representations $(G,\pi,\vec{d},\gamma)$, see
Definition~\ref{def:pirep}. Moreover, for simplicity let us assume
that the permutation action of $\img(\pi)$ is transitive, that is,
forms only a single orbit $\{1,\dots,n\}$. Note that this implies that
the dimension vector $\vec{d}=(d,\dots,d)$ is constant. By decomposing
an arbitrary $\pi$-representation into a direct sum we can always
reduce to the single-orbit case.

Now, $G$ acts on the index set $\{1,\dots,n\}$ via $\pi:G\to P$. Some
of the group elements of $G$ will leave $1\in\{1,\dots,n\}$
invariant. Let us denote this stabilizer by
\begin{equation}
  G_1 \eqdef 
  \Big\{ 
  g\in G
  ~\Big|~
  \pi(g)(1) = 1
  \Big\}
  .
\end{equation}
The restriction of the first block $\gamma_1$ of $\gamma$ to $G_1$ is
an actual representation of $G_1$, as this subgroup does not permute
it.

One can recover the whole representation matrix $\gamma$ from
$\gamma_1|_{G_1}$ as follows. First, fix a choice of group elements
$g_1\eqdef 1$, $g_i\in G$, $i=2,\dots,n$, such that
$\pi(g_i)(1)=i$. By the assumption of $P=\pi(G)$ having only a single
orbit, we can always find such $\{g_1, g_2, \dots, g_n\}$. This allows
us to factorize any group element into
\begin{equation}
  \forall g\in G
  ,~
  \forall 1\leq i \leq n
  \quad
  \exists h\in G_1: 
  \quad
  g = g_{\pi(g)(i)} \circ h \circ g_i^{-1}
\end{equation}
Due to the choice $g_1\eqdef 1$ the representation matrix
$\gamma_1(g_1)=\mathbf{1}_{d\times d}$. Since $g_i$, $i=2,\dots,n$
maps the first block to the $i$-th block, we can choose coordinates on
the $i$-th block such that 
\begin{equation}
  \gamma_1(g_i)=\mathbf{1}_{d\times d} 
  \quad
  \forall i=1,\dots,n
  .
\end{equation}
Using eq.~\eqref{eq:gammadef}, we can expand any group representation
matrix as
\begin{equation}
  \begin{split}
    \gamma(g) =&\; \gamma(g_{\pi(g)(i)}) \circ \gamma(h) \circ \gamma(g_i^{-1})
    \\
    =&\; 
    \mathbf{P}\big(\pi(g_{\pi(g)(i)}), \vec{d}\big)
    \diag\big( \gamma_1(g_{\pi(g)(i)}),~ \dots,~ \gamma_n(g_{\pi(g)(i)})\big)
    \\&\;
    \mathbf{P}\big(\pi(h), \vec{d}\big)
    \diag\big( \gamma_1(h),~ \dots,~ \gamma_n(h)\big)
    \\&\;
    \diag\big( \gamma_1(g_i)^{-1},~ \dots,~ \gamma_n(g_i)^{-1}\big)
    \mathbf{P}\big(\pi(g_i)^{-1}, \vec{d}\big)
  \end{split}
\end{equation}
Evaluating the permutation matrices, we see that the $i$-th block of
$\gamma(g)$ is
\begin{equation}
  \gamma_i(g) =
  \gamma_1(g_{\pi(g)(i)})
  \gamma_1(h)
  \gamma_1(g_i)^{-1}
  = 
  \gamma_1(h)
  \quad
  \forall i=1,\dots,n
  .
\end{equation}
Hence, $\gamma_1'=\gamma_1|_{G_1}$ determines the whole
$\pi$-representation $\gamma$.
\begin{example}[Induction]
  \label{ex:ind}
  Let $Q_8=\{\pm 1,\pm i, \pm j, \pm i j\}$ and $\pi(i) = (1,2)$,
  $\pi(j) = ()$.
  % \begin{equation}
  %   \pi(i) = (1,2)
  %   ,\quad
  %   \pi(j) = ()
  %   .
  % \end{equation}
  Then the stabilizer $(Q_8)_1 = \{j^\ell | \ell=0,\dots,3\} \simeq
  \Z_4$. Pick the representation
  \begin{equation}
    \gamma_1': (Q_8)_1 \to \C^\times
    ,\quad
    \gamma_1'(j^\ell) = \exp\left( \frac{2\pi i \ell}{4} \right)
  \end{equation}
  Now, let us choose $g_1=1$, $g_2 = i$. The $\pi$-representation
  $\big(G,\pi,(1,1),\gamma)$ thus generated is given by
  \begin{equation}
    \gamma(i) = 
    \begin{pmatrix}
      0 & -1 \\ 1 & 0
    \end{pmatrix}
    ,\quad
    \gamma(j) = 
    \begin{pmatrix}
      i & 0 \\ 0 & -i
    \end{pmatrix}
    .
  \end{equation}
\end{example}
This construction that is called \emph{induction}. It takes a
representation $\gamma_1':G_1 \to GL(d,\C)$ of a subgroup $G_1\subset
G$ and constructs a larger representation 
\begin{equation}
  \gamma = \Ind_{G_1}^G(\gamma_1')
  :~ 
  G\to GL\left(\frac{d |G|}{|G_1|},\C \right) 
  .
\end{equation}
To summarize, we have shown
\begin{theorem}[Defining data of a $\pi$-representation]
  \label{thm:pirepdata}
  A linear $\pi$-representation $(G,\pi,\vec{d},\gamma)$ such that
  $\img(\pi)$ has a single orbit is, up to linear coordinate changes,
  uniquely determined by 
  \begin{itemize}
  \item The permutation $P$ acting on $\{1,\dots,n\}$,
  \item a group homomorphism $\pi:G\to P$,
  \item the dimension $d\in\Z$ of a single block, and
  \item a linear representation $\gamma_1':G_1\to GL(d)$.
  \end{itemize}
  The corresponding $\pi$-representation is then
  \begin{equation}
    \Big( 
    G,~ 
    \pi,~ 
    \underbrace{(d,\dots,d)}_{n},~ 
    \Ind_{G_1}^G( \gamma_1' ) \Big) 
    .
  \end{equation}
\end{theorem}
Finally, note that there is an inner product on the group characters,
\begin{equation}
  (\chi,\psi) = 
  \frac{1}{|G|} 
  \sum_{g\in G} \chi(g) \overline{\psi(g)}
  ~
  \in \Z
  .
\end{equation}
With respect to this inner product, induction and
restriction\footnote{Restriction is just the ordinary pullback
  $\Res^G_H(\chi)\eqdef \chi|_{H}:H\to\C^\times$ of a character
  $\chi:G\to\C^\times$ to a subgroup $H\subset G$.} are adjoint
functors. That is, given a subgroup $H\subset G$ and characters $\chi$
of $H$ and $\psi$ of $G$,
\begin{equation}
  \Big<
    \Ind_H^G(\chi)
    ,~
    \psi
  \Big>
  =
  \Big<
    \chi
    ,~
    \Res^G_H(\psi)
  \Big>
  .
\end{equation}
Therefore, the character of an induced representation can be computed
without explicitly constructing the induced representation.

\subsection{Generalized Schur Covers}
\label{sec:generalizedschur}

Similar to the usual case of projective representations, we can turn
projective representations into linear representations by enlarging
the group. The basic recipe is the same as in
\autoref{sec:projective}: Given a projective representation
$\gamma:G\to \prod_i PGL(d_i)$, we can pick generators $g_1$, $\dots$,
$g_k$ of $G$ and matrices $\gamma(g_i)\in \prod_i GL(d_i)$ that
generate $\gamma$ projectively. As a matrix group, the $\gamma(g_i)$
generate a potentially larger group 
\begin{equation}
  \Gt \eqdef \left< \gamma(g_1),~\dots,~\gamma(g_k) \right>
\end{equation}
which maps onto $G$ in the tautological way $\Gt\to G$, $\gamma(g_i)
\mapsto g_i$. 

However, there are some differences. Most notably, the short exact
sequence
\begin{equation}
  \label{eq:GeneralSES}
  1
  \longrightarrow
  K
  \longrightarrow
  \Gt
  \longrightarrow
  G
  \longrightarrow
  1
\end{equation}
is no longer a central extension; In fact, the kernel $K\subset \Gt$
not only consists of matrices proportional to the identity matrix, but
also of the form $\bigoplus \zeta_i \mathbf{1}_{d_i\times d_i}$ with
not all $\zeta_i\in \C^\times$ being equal. Nevertheless, the
induction construction reviewed in \autoref{sec:ind} still works: A
projective representation of the stabilizer $G_1$ determines a twisted
representation of its ordinary Schur cover $\Gt_1$, which induces a
multi-twisted\footnote{We call a representation $\gamma:G\to\prod
  PGL(d_i)$ multi-projective. Lifting it to a linear representation
  yields a (non-unique) multi-twisted representation
  $\tilde\gamma:G\to \prod GL(d_i)$.} representation of $\Gt$
corresponding to a multi-projective representation of $G$. That way,
we can find a finite cover $\Gt$ for each finite group. However, $\Gt$
can be strictly larger than the ordinary Schur cover:
\begin{example}[A generalized Schur cover]
  Consider the group $G=\Z_4\times \Z_4=\{ (a,b) | 0\leq a,b\leq 3\}$
  acting on CICY \#21 via
  \begin{equation}
    \renewcommand{\arraystretch}{1.7}
    \begin{tabular}[c]{c|c|c|} 
      \multicolumn{1}{c}{} & 
      \multicolumn{1}{c}{{$\big( p_1 \big)$}} &
      \multicolumn{1}{c}{{$\big( p_2 \big)$}} \\
      \cline{2-3} 
      {$\CP_1=\CP^{2}$} & 
      $1$ & $1$
      \\ \cline{2-3}
      {$\CP_2=\CP^{2}$} & 
      $0$ & $2$
      \\ \cline{2-3}
      {$\CP_3=\CP^{2}$} & 
      $2$ & $0$
      \\ \cline{2-3}
      {$\CP_4=\CP^{2}$} & 
      $0$ & $2$
      \\ \cline{2-3}
      {$\CP_5=\CP^{2}$} & 
      $2$ & $0$
      \\ \cline{2-3}
    \end{tabular}
    \eqdef 
    C
    \;,\qquad
    \begin{array}{r@{\;=\;}l}
      \pi_r(1,0) & (2,3)(4,5) 
      \\
      \pi_r(0,1) & (2,4)(3,5) 
      \\
      \pi_c(1,0) & (1,2)
      \\
      \pi_c(0,1) & ()
      .
    \end{array}
  \end{equation}
  This defines the CICY group $(C,G,\pi_r,\pi_c)$. 
  
  A freely acting projective CICY group action is
  $(C,G,\pi_r,\gamma,\pi_c,\rho)$ with the representation matrices
  \begin{equation}
    \begin{aligned}
      \gamma(1,0) =&\;
      \begin{pmatrix}
        0 & 1 \\ 
        1 & 0
      \end{pmatrix}
      \oplus
      \left(\begin{smallmatrix}
          0 & 0 & 1 & 0 & 0 & 0 & 0 & 0 \\
          0 & 0 & 0 & 1 & 0 & 0 & 0 & 0 \\
          0 & -i & 0 & 0 & 0 & 0 & 0 & 0 \\ 
          i & 0 & 0 & 0 & 0 & 0 & 0 & 0 \\
          0 & 0 & 0 & 0 & 0 & 0 & i & 0 \\
          0 & 0 & 0 & 0 & 0 & 0 & 0 & i \\
          0 & 0 & 0 & 0 & 0 & 1 & 0 & 0 \\
          0 & 0 & 0 & 0 & -1 & 0 & 0 & 0 
        \end{smallmatrix}\right)
      , \quad&
      \gamma(0,1) =&\; 
      \begin{pmatrix}
        i & 0 \\
        0 & -i
      \end{pmatrix}
      \oplus
      \left(\begin{smallmatrix}
          0 & 0 & 0 & 0 & 1 & 0 & 0 & 0 \\ 
          0 & 0 & 0 & 0 & 0 & 1 & 0 & 0 \\ 
          0 & 0 & 0 & 0 & 0 & 0 & 1 & 0 \\ 
          0 & 0 & 0 & 0 & 0 & 0 & 0 & 1 \\ 
          -i & 0 & 0 & 0 & 0 & 0 & 0 & 0 \\ 
          0 & i & 0 & 0 & 0 & 0 & 0 & 0 \\ 
          0 & 0 & i & 0 & 0 & 0 & 0 & 0 \\ 
          0 & 0 & 0 & -i & 0 & 0 & 0 & 0
        \end{smallmatrix}\right)
      ,
      \\
      \rho(1,0) =&\; 
      \begin{pmatrix}
        0 & 1 \\
        1 & 0
      \end{pmatrix}
      ,\quad&
      \rho(0,1) =&\;
      \begin{pmatrix}
        -i & 0 \\ 
        0 & i
      \end{pmatrix}
      .
    \end{aligned}
  \end{equation}
  A basis for the $3$-dimensional space of invariant homogeneous
  polynomials is
  \begin{equation}
    \begin{array}{r@{\;\eqdef\;\big(~}c@{,~}c@{~\big)}}
      \vec{p}^{(1)} &
      z_2 z_5 z_6 z_9 z_{10} &
      -z_1 z_3 z_4 z_7 z_8 
      \\
      \vec{p}^{(2)} &
      z_1 z_5^2 z_{10}^2+z_1 z_6^2 z_9^2 & 
      -z_2 z_3^2 z_8^2-z_2 z_4^2 z_7^2 
      \\
      \vec{p}^{(3)} &
      z_1 z_5^2 z_9^2+z_1 z_6^2 z_{10}^2 &
      -z_2 z_3^2 z_7^2-z_2 z_4^2 z_8^2 
    \end{array}
    ,
  \end{equation}
  and one can show that a generic linear combination defines a
  fixed-point free smooth Calabi-Yau threefold.

  Clearly, $|G|=16$. A Schur cover, that is, a smallest group that
  linearizes any projective $G$-representation, is the Heisenberg
  group $\Z_4 \ltimes (\Z_4\times\Z_4)$ and has $64$ elements. This
  group is also sufficient to linearize the column
  $\pi$-representations.  However, it is insufficient to linearize the
  row $\pi$-representation $(G,\pi_r,(2,2,2,2,2),\gamma)$. The
  matrices $\gamma(1,0)$, $\gamma(0,1)$ generate a matrix group of
  order $256$.  Linearizing every row and column $\pi$-representation
  simultaneously requires a covering group of order $512$.
\end{example}

\section{Character-Valued Indices}
\label{sec:indices}

\subsection{Invariant and Equivariant Line Bundles}
\label{sec:linebundle}

Consider a line bundle $\Lsheaf$ on a complex manifold $X$ with a
group $G$ acting on $X$. Although we are primarily interested in free
actions, we will also consider group actions with fixed points for the
purposes of this subsection. 

The line bundle $\Lsheaf$ is invariant if $g^\ast \Lsheaf \simeq
\Lsheaf$ for all $g\in G$. If $\Pic^0(X)=1$, as is the case for proper
Calabi-Yau threefolds, the line bundles are classified by their first
Chern class. In that case $\Lsheaf$ is invariant if and only if
\begin{equation}
  c_1(\Lsheaf) \in H^2(X,\Z)^G
  .
\end{equation}
Each isomorphism $g^\ast \Lsheaf \simeq \Lsheaf$ defines a linear map
\begin{equation}
  \gamma(g): H^0(X,\Lsheaf) \longrightarrow  H^0(X,\Lsheaf)
  .
\end{equation}
However, the linear maps $\gamma(g)$ need not be a group homomorphism,
that is, $\gamma(g)\gamma(h)\not=\gamma(gh)$. Therefore, the
representation matrices $\gamma(g)$ generate a covering group $\Gt$
with kernel $K$, 
\begin{equation}
  1
  \longrightarrow
  K
  \longrightarrow
  \Gt
  \longrightarrow
  G
  \longrightarrow
  1
  .
\end{equation}
In the case where $X = \prod \CP^{d_i}$ is the ambient space of a
product of projective spaces, the short exact sequence is of course
identical to eq.~\eqref{eq:GeneralSES}.

A line bundle is \emph{equivariant} if it is invariant and the
representation matrices \emph{do} form a representation of the group
$G$ acting on the base space. Note that
\begin{itemize}
\item Not every $G$-invariant line bundle is $G$-equivariant.
\item Every $G$-invariant line bundle is $\Gt$-equivariant for some
  sufficient extension $\Gt\to G$. The kernel $K$ acts trivially on
  the base space $X$.
\item Every $G$-invariant line bundle is $\Z_k$-equivariant for every
  cyclic subgroup $\Z_k\subset G$.
\end{itemize}

\subsection{Implications of Freeness}
\label{sec:free}

Recall the generalization of the Lefshetz fixed point theorem to
holomorphic vector bundles~\cite{0151.31801}: Given a bundle $\Vsheaf$
over $X$ and a holomorphic map $f:X\to X$ with isolated\footnote{The
  case of non-isolated fixed points is essentially
  similar~\cite{MR0236951}. We only restrict to isolated fixed points
  for ease of presentation.} fixed points together with an isomorphism
$F:f^\ast \Vsheaf\to \Vsheaf$. Then this implies an action on the
bundle-valued cohomology groups via the double pull-back
\begin{equation}
  \vcenter{\xymatrix{
      &
      H^i(X,f^\ast\Vsheaf)
      \ar^-{F^\ast}[dr]
      \\
      \mathllap{H(f,F):\quad}
      H^i(X,\Vsheaf) \ar[rr]  
      \ar^-{f^\ast}[ur]
      & & 
      H^i(X,\Vsheaf)
      .
  }}
\end{equation}
Like the vector spaces $H^i(X,\Vsheaf)$, this map can depend on
moduli. However, the Euler characteristic
\begin{equation}
  \chi(f,F) 
  \eqdef
  \sum_i (-1)^i \Tr H^i(f,F)
  = 
  \sum_{P \in X^f}
  \frac{\Tr F_P}{\det\big( 1- df_P \big)}
  .
\end{equation}
is invariant under deformations and can be computed from data
localized at the fixed point set $X^f$ alone.

We always defined group actions on CICY manifolds $X$ via linear
$\pi$-representation $(\Gt,\pi_r,\vec{d},\gamma)$. Clearly, this
defines maps $\gamma(g):X\to X$. Moreover, by not only defining the
projective action but also the linearized action on the homogeneous
coordinates, we implicitly define isomorphisms $\gamma(k)^\ast(
\Lsheaf )\to \Lsheaf$ on any $G$-invariant holomorphic line bundle
$\Lsheaf$. Therefore, we have a well-defined action of
$(\Gt,\pi_r,\vec{d},\gamma)$ on the bundle cohomology groups
$H^i(X,\Lsheaf)$. By setting
\begin{equation}
  \chi(\Lsheaf)(g) 
  \eqdef
  \sum_i (-1)^i \Tr_{H^i(X,\Lsheaf)}\big( \gamma(g)^\ast \big)
  \qquad 
  \forall g\in \Gt
\end{equation}
we can extend the holomorphic Euler characteristic to a
one-dimensional representation of $\Gt$. Clearly, evaluating at $1\in
\Gt$ simplifies to the usual holomorphic Euler characteristic. Using
the fixed point theorem, we conclude that if $\Lsheaf$ is
$G$-invariant $(\Rightarrow$ $\Gt$-equivariant) and $g\in \Gt$ acts
freely on $X$, then $\chi(\Lsheaf)(g)=0$. 

If $\Lsheaf$ is already $G$-equivariant and $G$ acts freely, then we
furthermore learn that $X/G$ is a smooth manifold with holomorphic
line bundle $\Lsheaf / \gamma$. In this case, $\chi(\Lsheaf)(1) =
|G|\; \chi(X/G, \Lsheaf/\gamma)$ must be divisible by the order $|G|$ of
the group.
\begin{definition}[Free type of a character]
  \label{def:freetype}
  Consider a $G$-action on a CICY $X$ defined by an extension
  \begin{equation}
    1 
    \longrightarrow
    K
    \longrightarrow
    \Gt
    \longrightarrow
    G 
    \longrightarrow
    1
  \end{equation}
  and a linear CICY group action $(C,\Gt,\pi_r,\gamma,\pi_c,\rho)$.
  We say that the character-valued index $\chi(\Lsheaf):\Gt\to
  \C^\times$ of a $G$-invariant holomorphic line bundle is of free
  type if 
  \begin{itemize}
  \item $\chi(\Lsheaf)(g)=0 \quad \forall g\in \Gt-K$, and 
  \item if $\Lsheaf$ is $G$-equivariant, then $\tfrac{1}{|G|}
    \chi(\Lsheaf)(1) \in \Z$.
  \end{itemize}
\end{definition}
Clearly, if the $G$-action is free then the index is always of free
type.

\subsection{(Anti-)Symmetrizations and Induction}
\label{sec:AltInd}

As we discussed in \autoref{sec:ind}, the induction extends the group
action on the homogeneous coordinates of a single projective space to
the permutation orbit. Although this unambiguously defines the group
action on the combined homogeneous coordinates, it is not quite what
we need to compute the cohomology of line bundles on the product of
projective spaces. 
\begin{example}[Induction vs. Cohomology]
  \label{ex:indcoh}
  Consider the permutation action as in Example~\ref{ex:ind}. Now, let
  us start with the representation 
  \begin{equation}
    \gamma_1': 
    (Q_8)_1 \to GL(3,\C)
    ,\quad
    \gamma_1'(j^\ell) = \diag\big(1,~i^\ell,~ (-1)^\ell \big)
    . 
  \end{equation}
  The induced $Q_8$-representation $\gamma =
  \Ind_{(Q_8)_1}^{Q_8}(\gamma_1')$ is
  \begin{equation}
    \gamma(i) = 
    \begin{pmatrix}
      0 &   0 &   0 &   1 &   0 &   0  \\
      0 &   0 &   0 &   0 &   1 &   0  \\
      0 &   0 &   0 &   0 &   0 &   1  \\
      1 &   0 &   0 &   0 &   0 &   0  \\
      0 &   1 &   0 &   0 &   0 &   0  \\
      0 &   0 &  -1 &   0 &   0 &   0 
    \end{pmatrix}
    ,\quad
    \gamma(j) = 
    \begin{pmatrix}
      1 &      0 &      0 &      0 &      0 &      0  \\
      0 &     -1 &      0 &      0 &      0 &      0  \\
      0 &      0 &      i &      0 &      0 &      0  \\
      0 &      0 &      0 &      1 &      0 &      0  \\
      0 &      0 &      0 &      0 &     -1 &      0  \\
      0 &      0 &      0 &      0 &      0 &     -i 
    \end{pmatrix}
    .
  \end{equation}
  Now, consider $(Q_8,\pi,(3,3),\gamma)$ as a $\pi$-representation
  acting on $\CP^2_{[x_0:x_1:x_2]}\times \CP^2_{[y_0:y_1:y_2]} =
  \CP_1\times \CP_2$. Using the standard identification between
  sections of $\Osheaf(1)$ and homogeneous coordinates, we identify
  the representations
  \begin{equation}
    \begin{split}
      \gamma_1' =&\;
      H^0\big( \CP_1, \Osheaf(1) \big)
      ,\\ 
      \gamma =&\;
      \Ind_{(Q_8)_1}^{Q_8}(\gamma_1') = 
      H^0\big( \CP_1, \Osheaf(1) \big)
      \oplus
      H^0\big( \CP_2, \Osheaf(1) \big)
      \\
      =&\;
      H^0\big( \CP_1\times \CP_2, 
      \Osheaf(1,0)\oplus \Osheaf(0,1) \big)
      = \Span_\C\big\{x_0,x_1,x_2, y_0,y_1,y_2\big\}
      .
    \end{split}
  \end{equation}
  But we would like to know the cohomology of an \emph{invariant} line
  bundle, for example
  \begin{equation}
    H^0\big( \CP_1\times \CP_2, \Osheaf(1,1) \big) 
    =
    H^0\big( \CP_1, \Osheaf(1) \big)
    \otimes
    H^0\big( \CP_2, \Osheaf(1) \big)
    =
    \Span_\C\big\{ x_i y_j \big| 0\leq i,j \leq 2 \big\}
    .
  \end{equation}
\end{example}

The problem is that the induction procedure $\Ind_H^G$ adds (as direct
sum $\oplus$) the $H$-representations in order to get the
$G$-representations, but for the purposes of computing the cohomology
groups of projective spaces we should multiply them (form the
symmetrized tensor product $\odot$). Hence, we are led to define a new
operation
\begin{definition}[$\SymInd$ and $\AltInd$]
  \label{def:SymAltInd}
  Let $H\subset G$ and $\gamma_1':H\to GL(n)$ a representation of
  $H$. We know that the induced representation is of the form
  eq.~\eqref{eq:gammadef}
  \begin{equation}
    \Ind_H^G(\gamma_1')(g)
    =
    \mathbf{P}\big(\pi(g), \vec{d}\big)
    ~
    \big( \gamma_1(g) \oplus \cdots \oplus \gamma_n(g)\big)
    .
  \end{equation}
  Let us define the associated operations
  \begin{equation}
    \begin{split}
      \SymInd_H^G(\gamma_1')(g) 
      =&\;
      \gamma_1(g) \odot \cdots \odot \gamma_n(g)
      ,\\
      \AltInd_H^G(\gamma_1')(g) 
      =&\;
      (-1)^{|\pi(g)|}
      \gamma_1(g) \odot \cdots \odot \gamma_n(g)
      ,
    \end{split}
  \end{equation}
  where $|\pi(g)|$ is the signature of the permutation $\pi(g)$. If
  the representation is $\Z_2$-graded, then we furthermore define
  \begin{equation}
    \GrInd_H^G(\gamma_1)
    =
    \begin{cases}
      \SymInd_H^G(\gamma_1') &
      \text{ if $\gamma_1'$ is even,}
      \\
      \AltInd_H^G(\gamma_1') &
      \text{ if $\gamma_1'$ is odd.}
    \end{cases}
  \end{equation}
\end{definition}
Clearly, this definition of $\SymInd$/$\AltInd$ does not refer to
specific coordinates and therefore extends to operations on group
characters. In \autoref{sec:SymIndChar}, we will present explicit
formulas that are necessary to efficiently compute the
character-valued indices that appear in the CICY group classification
algorithm.

Let us further note that the definition of $\SymInd$ is exactly what
is needed to compute the cohomology groups of line bundles on products
of projective spaces:
\begin{example}[Continuation of Example~\ref{ex:indcoh}]
  \begin{equation}
    \renewcommand{\arraystretch}{1.7}
    \begin{array}{|r|ccccc|}
      \hline
      \hbox{\rm Conj.classes}(Q_8) & 1 & i & j & -1 & ij \\
      \hline\hline
      \Ind(\gamma_1')    &  6 & 0 & 0 & 2 & 0 \\
      \SymInd(\gamma_1') &  9 & 1 & 1 & 1 & 1 \\
      \AltInd(\gamma_1') & 9 & -1 & 1 & 1 & -1 \\
      \hline
    \end{array}
  \end{equation}
  The cohomology of the line bundle $\Osheaf(1,1)$ is\footnote{In the
    context of $G$-manifolds and $G$-equivariant vector bundles, we
    write $H^\bullet(\cdots)$ for the $G$-representation on the
    cohomology and $h^\bullet(\cdots)$ for the corresponding
    $G$-character.}
  \begin{equation}
    h^0\big( \CP_1\times \CP_2, \Osheaf(1,1) \big) =
    \SymInd_{(Q_8)_1}^{Q_8}(\gamma_1')
  \end{equation}
  as a $Q_8$-character. Note that $\gamma_1'$ and the permutation
  action are precisely the defining data for the $\pi$-representation,
  see Theorem~\ref{thm:pirepdata}.
\end{example}

If we have a general $\pi$-representation $(G,\pi_r,\vec{d},\gamma)$
acting on $\prod_{k=1}^n \CP_k = \prod \CP^{d_k}$, then we have to
split the product into $\pi_r$-orbits and apply the $\SymInd$
construction to each orbit. Let us define the index set and its
$\pi_r$-orbits to be
\begin{equation}
  S_n \eqdef 
  \{1,\dots,n\} = 
  \{1,\dots\} \cup \cdots \cup \{\dots, n\} 
  = 
  \bigcup_{G\{i\} \in S_n/G } G\{i\}
  .
\end{equation}
By abuse of notation, we denote by $i$ also the embedding of the
$i$-th factor $\CP_i$ in the product,
\begin{equation}
  i:~
  \CP_i 
  \longrightarrow
  \prod_{k=1}^n \CP_k
  .
\end{equation}
Finally, note that exchanging two odd-degree cohomology groups incurs
an extra minus sign. Therefore, the character-valued cohomology of a
$G$-equivariant line bundle $\Lsheaf$ is
\begin{equation}
  h^\bullet
  \Big( \prod_{k=1}^n \CP_k,~ 
  \Lsheaf \Big)
  =
  \prod_{G\{i\} \in S_n/G }
  \GrInd_{G_i}^G \Big(
  h^\bullet\big( \CP_i,~ i^\ast \Lsheaf \big)
  \Big),
\end{equation}
where $\GrInd$ is symmetric or anti-symmetric depending on the mod-$2$
cohomological degree of $h^\bullet\big( \CP_i,~ i^\ast \Lsheaf
\big)$. The corresponding character-valued Euler characteristic is 
\begin{equation}
  \label{eq:CohPnGrInd}
  \chi
  \Big( \prod_{k=1}^n \CP_k,~ 
  \Lsheaf \Big)
  =
  \sum_{\vec{q}} 
  (-1)^{\sum_{G\{i\} \in S_n/G } [G:G_i] q_i}
  \prod_{G\{i\} \in S_n/G }
  \GrInd_{G_i}^G \Big(
  h^{q_i}\big( \CP_i,~ i^\ast \Lsheaf \big)
  \Big),  
\end{equation}
where the summation over all possible degree vectors $\vec{q}\in
\Z^{|S_n/G|}$ has, of course, only finitely many non-zero summands.

\subsection{The Koszul Spectral Sequence}
\label{sec:Koszul}

Consider a complete intersection cut out by $m$ transverse
polynomials. Each polynomial equation $p_i=0$ defines a divisor 
\begin{equation}
  D_j \eqdef \Big\{ p_j=0 \Big\}
  ~\subset~
  \prod_{i=1}^n \CP_i
  .
\end{equation}
An immediate consequence of a complete intersection $X \subset \prod
\CP_i$ is that we have a Koszul resolution\footnote{By $\Osheaf$ we
  will always denote the trivial line bundle on the \emph{ambient}
  space $\prod \CP_i$.}
\begin{equation}
  \vcenter{\xymatrix{
      0 
      \ar[r]
      & 
      \Osheaf\big( -\sum D_j \big)
      \ar|-{\textstyle ~\cdots~}[rr]
      &&
      \displaystyle
      \smash{\bigoplus_{j< k}}
      \Osheaf(-D_j-D_k)
      \ar[r]
      &
      \bigoplus \Osheaf(-D_j)
      \ar[r]
      &
      \underline{ 
        \Osheaf
      }
      \ar[r]
      &
      0
      .
    }}
\end{equation}
That is, the above sequence is exact everywhere except at the
underlined entry. At that position, the cohomology is $\Osheaf_X$. In
other words, the Koszul complex is equivalent to $\Osheaf_X$ in the
derived category, and we can interchange them for the purposes of
computing bundle cohomology. After tensoring with a line bundle
$\Lsheaf$, the associated hypercohomology spectral sequence reads
\begin{equation}
  \label{eq:KoszulSS}
  E_1^{-p,q} = 
  H^q\Big(\prod \CP_i, 
  \textstyle
  \bigoplus_{1\leq j_1<\cdots<j_p\leq m}\Osheaf(-D_{j_1}-\cdots -
  D_{j_p})
  \otimes \Lsheaf
  \Big)
  ~
  \Rightarrow
  ~
  H^{-p+q}\big( X, \Lsheaf|_X \big)
\end{equation}
Note that all non-vanishing entries are in the second quadrant. To
evaluate all the higher differentials in the spectral sequence is, of
course, a lot of work. However, any non-trivial differential removes
the same subspace from the even and from the odd cohomology groups,
leaving the Euler characteristic invariant. Therefore, we can compute
the character-valued index already from the $E_1$-tableau by
pretending that all higher differentials vanish. One obtains
\begin{equation}
  \chi(X,\Lsheaf|_X) 
  = 
  \sum_{1\leq j_1<\cdots<j_p\leq m} (-1)^p 
  \chi\Big( 
  \Osheaf(-D_{j_1}-\cdots -D_{j_p})\otimes \Lsheaf 
  \Big)
  =
  \sum_{p,q} (-1)^{p+q} E_1^{-p,q} 
\end{equation}
A good way of dealing with the indices $1\leq j_1<\cdots<j_p\leq n$ in
the resolution is to consider them as basis elements of the (formal)
exterior algebra generated by the polynomials $p_{j_1} \wedge \cdots
\wedge p_{j_p}$. 
\begin{example}[Koszul resolution]
  \label{ex:Koszul}
  By abbreviating $\Osheaf(-D_{j_1}-\cdots
  -D_{j_p})=\Osheaf_{j_1\wedge \cdots\wedge j_p}$ we can write the
  Koszul complex for $r=3$ transverse polynomials as
  \begin{equation}
    \vcenter{\xymatrix{
        0 
        \ar[r]
        & 
        \Osheaf_{1\wedge 2\wedge 3}
        \ar^-{(p_3,p_2,p_1)}[r]
        &
        \Osheaf_{1\wedge 2}
        \oplus
        \Osheaf_{1\wedge 3}
        \oplus
        \Osheaf_{2\wedge 3}
        \ar^-{
          \left(\begin{smallmatrix}
              p_2 & -p_1 & 0 \\
              -p_3 & 0 & p_1 \\
              0 & p_3 & -p_2 \\
            \end{smallmatrix}\right)
        }[rr]
        &&
        \Osheaf_1
        \oplus
        \Osheaf_2
        \oplus
        \Osheaf_3
        \ar^-{
          \left(\begin{smallmatrix}
              p_1 \\ p_2 \\ p_3
            \end{smallmatrix}\right)
        }[r]
        &
        \underline{ 
          \Osheaf
        }
        \ar[r]
        &
        0
        .
      }}
  \end{equation}  
\end{example}

\subsection{Equivariant Koszul}
\label{sec:KoszulEquiv}

\subsubsection{No Permutations}

First, let us assume that there are no permutations, but only a linear
$G$-action on each projective space and each polynomial. Then we can
easily compute the cohomology of each line bundle $H^q(\prod\CP_i,
\Osheaf_{j_1\wedge \cdots\wedge j_p})$ as a $G$-representation, using
the notation of Example~\ref{ex:Koszul}. However, if the polynomial
equations are not $G$-invariant, then the index must depend on their
transformation as well! Following the maps through the Koszul
resolution until we end up at the homological degree-$0$ piece, we see
that $\Osheaf_{j_1\wedge \cdots\wedge j_p}$ ends up being multiplied
by $p_{j_1}$, $p_{j_2}$, $\dots$, $p_{j_p}$. Therefore, the
its contribution to the character-valued index must be $p_{j_1}\cdots
p_{j_p}\chi(\Osheaf_{j_1\wedge\cdots\wedge j_p})$, where we consider
the polynomials as $G$-characters.

\subsubsection{With Permutations}

This gets more complicated when we consider the case where the
$G$-action permutes the polynomials by a permutation action $\pi:G\to
P$. Since the polynomials appear with different signs in the maps of
the Koszul resolution, permuting them yields an extra minus sign
corresponding to the signature of the permutation. Therefore, the
contribution to the character-valued index is
\begin{equation}
  \begin{split}
    \chi(\Lsheaf|_X) =&\;
    \sum (-1)^p
    ~
    p_{j_1} \wedge \cdots \wedge p_{j_p}
    ~
    \chi\Big( \Osheaf_{j_1\wedge\cdots\wedge j_p} \otimes \Lsheaf\Big)
    \\
    =&\;
    \sum_{\wedge \vec{\jmath} \in \Lambda_m} (-1)^p
    ~
    p_{\wedge\vec{\jmath}}
    ~
    \chi\Big( \Osheaf_{\wedge\vec{\jmath}} \otimes \Lsheaf\Big)
    ,
  \end{split}
\end{equation}
where we used the notation 
\begin{equation}
  \Lambda_m
  \eqdef
  \Big\{ 
  j_1 \wedge\cdots\wedge  j_p
  ~\Big|~
  0\leq p \leq m
  ,~ 
  1\leq j_1 < \cdots < j_p \leq m
  \Big\}  
\end{equation}
for the standard basis of anti-symmetrized indices and 
\begin{equation}
  p_{\wedge \vec{\jmath}} = 
  p_{\wedge (j_1,\dots,j_p)} 
  \eqdef 
  p_{j_1} \wedge \cdots \wedge p_{j_p}
\end{equation}
for the exterior powers of the polynomials thought of as group
characters. However, the above equation for $\chi(\Lsheaf|_X)$ is only
useful if the multi-index $\wedge \vec{\jmath} = j_1\wedge\cdots\wedge
j_p$ is invariant under the permutation action; Otherwise, the group
action will exchange different summands and we still do not have a
closed expression for the index.

To write a general equation, we have to decompose the multi-indices
into orbits of the permutation action and choose representatives
\begin{equation}
  \Lambda_r / G 
  \eqdef
  \Big\{ 
  [\wedge \vec{\jmath}_{(1)}],~  
  [\wedge \vec{\jmath}_{(2)}],~ 
  \dots \Big\}
  = 
  \Big\{ 
  \pm j_1 \wedge\cdots\wedge  j_p
  ~\Big|~
  0\leq p \leq n
  \Big\}
  \Big/ \big<\pm, G\big>
  .
\end{equation}
Each bundle $\Osheaf_{\wedge \vec{\jmath}}$ is then fixed under
\begin{equation}
  G_{\wedge \vec{\jmath}} = \Stab_{\wedge\vec{\jmath}} (G) = 
  \Big\{ g\in G ~\Big|~ 
  \pi(g)(\wedge \vec{\jmath}) = \pm \wedge \vec{\jmath} \Big\}
  ,
\end{equation}
and, therefore, 
\begin{equation}
  \chi\big( \Osheaf_{\wedge \vec{\jmath}} \otimes \Lsheaf \big)
  :~
  G_{\wedge \vec{\jmath}}
  \longrightarrow
  \C^\times
\end{equation}
is a character of the stabilizer. 

The other $G_{\wedge \vec{\jmath}}\;$-character that enters the index
formula is $p_{\wedge \vec{\jmath}}$.  However, each individual
polynomial $p_j$ is a character of its stabilizer $G_j$ which, in
general, neither contains nor is contained in $G_{\wedge
  \vec{\jmath}}$. To proceed further, we have to decompose the
$G$-invariant index sets into $G$-orbits of a single index,
\begin{equation}
  \wedge \vec{\jmath} = 
  j_1 \wedge \cdots \wedge j_p 
  =
  \big(
  j_1 \wedge \cdots
  \big)
  \wedge
  \big(
  \cdots
  \big)
  \wedge \cdots \wedge
  \big(
  \cdots \wedge j_p
  \big)  
  = 
  \bigwedge_{j\in \vec{\jmath}/G}
  \Big( \wedge G (j) \Big)
\end{equation}
Now, consider the orbit $G(j)$ generated by $j$. To compute the
$p_{\wedge G(j)}$ as a character of $G_{\wedge \vec{\jmath}}$ we only
need knowledge of one of the polynomials (say, $p_j$) and the
permutation action of the group. One obtains that
\begin{equation}
  p_{\wedge \vec{\jmath}} = 
  \prod_{j\in \vec{\jmath}/G}
  \AltInd
  _{G_j \cap G_{\wedge \vec{\jmath}}}
  ^{G_{\wedge \vec{\jmath}}}
  \Big(
  \Res^{G_j}_{G_j \cap G_{\wedge \vec{\jmath}}}(p_j)
  \Big)
\end{equation}
as a character of $G_{\wedge \vec{\jmath}}$.

Finally, summing over the $\Lambda_m/G$-orbits and keeping
track of how the permutation acts on the summands is nothing but the
induction from the stabilizer $G_{\wedge \vec{\jmath}}$ to the full
group $G$. Therefore, we can write a closed expression for the
character-valued index as
\begin{equation}
  \chi(\Lsheaf|_X) = 
  \sum_{\wedge \vec{\jmath}\in \Lambda_m/G}
  (-1)^{|\wedge \vec{\jmath}|}
  \Ind_{G_{\wedge \vec{\jmath}}}^G
  \Big(
    p_{\wedge\vec{\jmath}} 
    ~
    \chi\big( 
    \Osheaf_{\wedge \vec{\jmath}} \otimes \Lsheaf 
    \big)
  \Big)
  .
\end{equation}

\subsubsection{General Case}

In the most general case, the group $G$ acts on the polynomials not
only via permutations, but also by forming non-trivial linear
combinations if the degrees allow for it. As in the CICY case, we
group polynomials of the same degree into vectors $\vec{p}_j$.
Moreover, we assign multiplicities $1\leq |j| \leq \dim(\vec{p}_j)$ to
each index, constant on permutation orbits, in order to keep track of
$|j|$-fold exterior powers $\wedge_{|j|}\vec{p}_j \eqdef
\vec{p}_j\wedge \cdots \wedge \vec{p}_j$ contributing to the
character-valued index. Here, the exterior powers are graded by
$|j|\tmod 2$.

Hence, the index set of interest is
\begin{equation}
  \Lambda_m
  \eqdef
  \Big\{ 
  i_1 \wedge\cdots\wedge  i_k
  ~\Big|~
  0\leq \textstyle \sum |j_\ell| \leq m
  ,~ 
  1\leq i_1<\cdots<i_k\leq m
  \Big\}
\end{equation}
The permutation action on the multi-indices-with-multiplicities can
then again be grouped into orbits
\begin{equation}
  \Lambda_m / G = 
  \bigcup_{\wedge \vec{\jmath}\in \Lambda_m/G}
  \Big\{ \wedge \vec{\jmath} \Big\}
  = 
  \bigcup_{\wedge \vec{\jmath}\in \Lambda_m/G}
  \Big\{ 
  \bigwedge_{j\in \vec{\jmath}/G}
  \big( \wedge G(j) \big)
  \Big\}
\end{equation}
Putting everything together, the closed form expression for the
character-valued index is
\begin{multline}
  \label{eq:koszulchar}
  \chi(\Lsheaf|_X) =
  \sum (-1)^{\sum |j_\ell|}
  ~
  \Big(\mathop{\wedge}_{|j_1|} p_{j_1}\Big) 
  \wedge \cdots \wedge 
  \Big(\mathop{\wedge}_{|j_k|} p_{j_k}\Big)
  ~
  \chi\Big( \Osheaf_{j_1\wedge\cdots\wedge j_k} \otimes \Lsheaf\Big)
  \\
  =
  \sum_{\wedge \vec{\jmath}\in \Lambda_m/G} (-1)^{|\wedge \vec{\jmath}\,|}
  \Ind_{G_{\wedge \vec{\jmath}}}^G
  \bigg[
  \chi\big( 
  \Osheaf_{\wedge \vec{\jmath}} \otimes \Lsheaf 
  \big)
  \prod_{j\in \vec{\jmath}/G}
  \GrInd
  _{G_j \cap G_{\wedge \vec{\jmath}}}
  ^{G_{\wedge \vec{\jmath}}}
  \Big(
  \bigwedge_{|j|}
  \Res^{G_j}_{G_j \cap G_{\wedge \vec{\jmath}}}(\vec{p}_j)
  \Big)    
  \bigg],
\end{multline}
where the grading in $\GrInd$ is $|j| \tmod 2$.

\subsection{Character-Valued Index}
\label{sec:CharIndex}

Let us now apply the Koszul resolution to the CICYs. Using
\eqref{eq:koszulchar}, the index of a line bundle on the Calabi-Yau
threefold is determined by the character-valued cohomology groups on
the ambient space and group theoretic information about the column
CICY group action. For each term in the resolution, we then apply
eq.~\eqref{eq:CohPnGrInd} in order to compute the cohomology groups on
the ambient space from the row CICY group action. We use the following
notation:
\begin{itemize}
\item
  \begin{math}
    \Gt_i = \Stab_{\{i\}}(\pi_r) = 
    \big\{ g \in \Gt ~\big|~ \pi_r(g)(j) = j \big\}
  \end{math} 
  is the stabilizer of the $i$-th row under the action of the row
  permutations.
\item 
  \begin{math}
    \Gt_j = \Stab_{\{j\}}(\pi_c) = 
    \big\{ g \in \Gt ~\big|~ \pi_c(g)(j) = j \big\}
  \end{math} 
  is the stabilizer of the $j$-th column under the action of the
  column permutations.
\item
  \begin{math}
    \Gt_{\wedge \vec{\jmath}} = \Stab_{\wedge \vec{\jmath}}(\pi_c) =
    \big\{ g \in \Gt ~\big|~ \pi_c(g)(\wedge \vec{\jmath}) = \pm
    \wedge \vec{\jmath} \big\}
  \end{math} 
  is the stabilizer of $\Osheaf_{\wedge \vec{\jmath}}$ in the Koszul
  resolution. 
\item The homogeneous coordinates of the $i$-th projective space
  $\CP_i$ form a (linear) representation of $\Gt_i$. Let us denote the
  restriction to the subgroup $\Gt_i \cap \Gt_{\wedge \vec{\jmath}}$
  by $\Res^{\Gt_i}_{\Gt_i \cap \Gt_{\wedge \vec{\jmath}}} (\CP_i)$.
\end{itemize}
The character-valued index of $\Lsheaf|_X$ on the Calabi-Yau threefold
$X$ is then
\begin{multline}
  \chi(\Lsheaf|_X)
  =
  \sum_{\wedge \vec{\jmath}~\in~ \Lambda_m/G 
    \vphantom{\Z^{|S_n / \Gt_{\wedge \vec{\jmath}}|}}} 
  \quad
  \sum_{\vec{q}~\in~ \Z^{|S_n / \Gt_{\wedge \vec{\jmath}}|}} 
  (-1)^{
    |\wedge \vec{\jmath}|
    + 
    \sum_{\Gt_{\wedge \vec{\jmath}}\{i\} \in 
      S_n/\Gt_{\wedge \vec{\jmath}} } [\Gt:\Gt_i] q_i
  }
  \\
  \Ind_{\Gt_{\wedge \vec{\jmath}}}^{\Gt}
  \Bigg\{
  \bigg[
  \prod_{\Gt_{\wedge \vec{\jmath}}\{i\} \in S_n/\Gt_{\wedge
      \vec{\jmath}} }
  \!\!\!
  \GrInd_{\Gt_i \cap \Gt_{\wedge \vec{\jmath}}}^{\Gt_{\wedge \vec{\jmath}}}
  h^{q_i}\Big(
  \Res^{G_i}_{G_i \cap G_{\wedge \vec{\jmath}}}  (\CP_i)
  ,~ 
  i^\ast (\Osheaf_{\wedge \vec{\jmath}} \otimes \Lsheaf)
  \Big)
  \bigg] 
  \\
  \times 
  \bigg[
  \prod_{j\in \vec{\jmath}/\Gt}
  \GrInd
  _{\Gt_j \cap \Gt_{\wedge \vec{\jmath}}}
  ^{\Gt_{\wedge \vec{\jmath}}}
  \Big(
  \bigwedge_{|j|}
  \Res^{\Gt_j}_{\Gt_j \cap \Gt_{\wedge \vec{\jmath}}}(\vec{p}_j)
  \Big)    
  \bigg]
  \Bigg\}
  .
\end{multline}
The importance of the above formula is that it expresses the index
using precisely the defining data of a CICY group action and only
group characters (instead of explicit representations).

\section{Calabi-Yau Groups}
\label{sec:CYgroups}

\begin{table}[htbp]
  \centering
  {
    \small
    \renewcommand{\arraystretch}{0.9}
    \begin{tabular}{|c|cc|ccl|}
  \hline
  Group & $|G|$ & ID & $N_\text{f}$ & $N_\text{f+s}$ & CICY \# \\
  \hline
  \hline
  $\Z_2$ & $2$ & $1$ & $166$ & $166$ & $\dots$  \\
  \hline
  $\Z_3$ & $3$ & $1$ & $31$ & $31$ & $\begin{minipage}{8.5cm}
    \vspace{2mm}
    \flushleft
    \scriptsize
    $6$, $14$, $18$, $26$, $242$, $536$, $1215$, $1306$, $2104$,
    $3388$, $3406$, $3413$, $3620$, $4415$, $5967$, $5982$, $6021$,
    $6024$, $6502$, $7206$, $7240$, $7246$, $7247$, $7300$, $7664$,
    $7669$, $7800$, $7808$, $7810$, $7878$, $7884$
    \vspace{2mm}
  \end{minipage}$  \\
  \hline
  $\Z_4$ & $4$ & $1$ & $23$ & $23$ & $\begin{minipage}{8.5cm}
    \vspace{2mm}
    \flushleft
    \scriptsize
$19$, $20$, $21$, $30$, $95$, $480$, $2564$, $2568$, $2572$,
$2639$, $5301$, $5452$, $6826$, $6836$, $6927$, $6947$, $7246$,
$7300$, $7484$, $7735$, $7745$, $7861$, $7862$
    \vspace{2mm}
  \end{minipage}$  \\
  $\Z_2 \times \Z_2$ & $4$ & $2$ & $40$ & $40$ & $\begin{minipage}{8.5cm}
    \flushleft
    \scriptsize
$15$, $19$, $20$, $21$, $22$, $480$, $2357$, $2534$, $2564$, $2566$,
$2568$, $2640$, $5256$, $5301$, $5302$, $5421$, $5452$, $6715$,
$6784$, $6788$, $6826$, $6828$, $6829$, $6836$, $6927$, $6947$,
$7435$, $7447$, $7462$, $7484$, $7487$, $7491$, $7522$, $7714$,
$7735$, $7745$, $7819$, $7823$, $7861$, $7862$
    \vspace{2mm}
  \end{minipage}$  \\
  \hline
  $\Z_5$ & $5$ & $1$ & $5$ & $5$ & $4335, 6655, 7447, 7761, 7890$  \\
  \hline
  $\Z_6$ & $6$ & $2$ & $4$ & $4$ & $6, 7206, 7246, 7300$  \\
  \hline
  $\Z_8$ & $8$ & $1$ & $7$ & $7$ & $\scriptstyle 19, 21, 2564, 6836, 6947, 7861, 7862$  \\
  $\Z_4 \times \Z_2$ & $8$ & $2$ & $11$ & $11$ & $\scriptstyle 19, 21, 2564, 2568, 6836, 6927, 6947, 7735, 7745, 7861, 7862$  \\
  $Q_8$ & $8$ & $4$ & $7$ & $7$ & $\scriptstyle 19, 21, 2564, 6836, 6947, 7861, 7862$  \\
  $\Z_2 \times \Z_2 \times \Z_2$ & $8$ & $5$ & $1$ & $1$ & $7861$  \\
  \hline
  $\Z_3 \times \Z_3$ & $9$ & $2$ & $6$ & $6$ & $14, 7240, 7669, 7808, 7878, 7884$  \\
  \hline
  $\Z_{10}$ & $10$ & $2$ & $3$ & $3$ & $4335, 7447, 7761$  \\
  \hline
  $\Z_3 \rtimes \Z_4$ & $12$ & $1$ & $2$ & $2$ & $7246, 7300$  \\
  $\Z_{12}$ & $12$ & $2$ & $2$ & $2$ & $7246, 7300$  \\
  \hline
  $\Z_4 \times \Z_4$ & $16$ & $2$ & $5$ & $3$ & $21, \hbox{\sout{6836}}, \hbox{\sout{6947}}, 7861, 7862$  \\
  $\Z_4 \rtimes \Z_4$ & $16$ & $4$ & $5$ & $5$ & $21, 6836, 6947, 7861, 7862$  \\
  $\Z_8 \times \Z_2$ & $16$ & $5$ & $5$ & $5$ & $21, 6836, 6947, 7861, 7862$  \\
  $\Z_8 \rtimes \Z_2$ & $16$ & $6$ & $2$ & $2$ & $21, 7862$  \\
  $\Z_4 \times \Z_2 \times \Z_2$ & $16$ & $10$ & $1$ & $1$ & $7861$  \\
  $\Z_2 \times Q_8$ & $16$ & $12$ & $5$ & $3$ & $21, \hbox{\sout{6836}}, \hbox{\sout{6947}}, 7861, 7862$  \\
  \hline
  $\Z_{10} \times \Z_2$ & $20$ & $5$ & $1$ & $1$ & $7447$  \\
  \hline
  $\Z_5 \times \Z_5$ & $25$ & $2$ & $2$ & $1$ & $7890, \hbox{\sout{7761}}$  \\
  \hline
  $(\Z_4 \times \Z_2) \rtimes \Z_4$ & $32$ & $2$ & $1$ & $1$ & $7861$  \\
  $\Z_8 \times \Z_4$ & $32$ & $3$ & $1$ & $1$ & $7861$  \\
  $\Z_8 \rtimes \Z_4$ & $32$ & $4$ & $1$ & $1$ & $7861$  \\
  $(\Z_8 \times \Z_2) \rtimes \Z_2$ & $32$ & $5$ & $1$ & $1$ & $7861$  \\
  $\Z_8 \rtimes \Z_4$ & $32$ & $13$ & $1$ & $1$ & $7861$  \\
  $\Z_4 \times \Z_4 \times \Z_2$ & $32$ & $21$ & $1$ & $1$ & $7861$  \\
  $\Z_2 \times (\Z_4 \rtimes \Z_4)$ & $32$ & $23$ & $1$ & $1$ & $7861$  \\
  $\Z_4 \rtimes Q_8$ & $32$ & $35$ & $1$ & $1$ & $7861$  \\
  $\Z_2 \times \Z_2 \times Q_8$ & $32$ & $47$ & $1$ & $1$ & $7861$  \\
  \hline
  $\Z_{10} \times \Z_5$ & $50$ & $5$ & $1$ & $0$ & $\hbox{\sout{7761}}$  \\
  \hline
  $\Z_8 \times \Z_8$ & $64$ & $2$ & $1$ & $0$ & $\hbox{\sout{7861}}$  \\
  $\Z_8 \rtimes \Z_8$ & $64$ & $3$ & $1$ & $0$ & $\hbox{\sout{7861}}$  \\
  $(\Z_4 \rtimes \Z_4) \rtimes \Z_4$ & $64$ & $68$ & $1$ & $0$ & $\hbox{\sout{7861}}$  \\
  $(\Z_2 \times Q_8) \rtimes \Z_4$ & $64$ & $72$ & $1$ & $0$ & $\hbox{\sout{7861}}$  \\
  $\Z_8 \rtimes Q_8$ & $64$ & $179$ & $1$ & $0$ & $\hbox{\sout{7861}}$  \\
  \hline
\end{tabular}
    
  }
  \caption{The free group actions on CICYs. The \sout{stricken out} numbers are
    non-smooth CICY. The ``ID'' field is the GAP
    \texttt{IdSmallGroup} of the
    group. $N_\text{f}$ is the number of CICY configurations admitting
    a free action,
    and $N_\text{f+s}$ is the number of \emph{smooth} CICY admitting a free
    action.
  }
  \label{tab:CICYgroups}
\end{table}
I ran the classification algorithm and found group actions allowed by
indices on $195$ CICY configurations. Usually, there is more than one
action of the same group for any given CICY configuration. It is
difficult to distinguish truly distinct actions from those that are
related by an automorphism of the manifold. For example, the two free
$\Z_3\times\Z_3$ actions on the CICY $\#19$ investigated
in~\cite{Braun:2004xv} and~\cite{Braun:2007tp, Braun:2007xh,
  Braun:2007vy, Triadophilia} yield quotients with different complex
structures, but are neither distinguished by topological invariants
like Betti numbers nor by Gromov-Witten invariants, at least not by
those that have been computed so far. With this caveat in mind, the
CICY configurations admitting free group actions are listed in
\autoref{tab:CICYgroups}. Note that, in a few cases indicated by a
stricken-out CICY number in the table, all linear combinations of
invariant polynomials fail to be transverse. These define free group
actions on singular CICY threefolds. Moreover, note that most
$2$-groups are realized on the CICY $\#7861$, the complete
intersection of $4$ quadrics in $\CP^7$. These were classified
previously\footnote{Note that the order-$32$ group
  $\Z_2\times(\Z_4\rtimes\Z_4)=\hbox{\texttt{SmallGroup(32,23)}}$ is
  omitted in~\cite{hua-2007}} in~\cite{BeauvilleNonAbel, MR2373582,
  hua-2007}.

\begin{table}[htbp]
  \centering
  {
    \small
    \renewcommand{\arraystretch}{0.9}
    \begin{tabular}{|cc|cc|cc|c@{~~}c@{~~}c@{~~}c@{~~}c@{~~}c@{~~}c@{~~}c@{~~}c@{~~}c|}
  \hline
  $|G|$ & ID & $N_\text{f}$ & $N_\text{f+s}$ & 
  \begin{sideways}Abelian invariants\end{sideways} &
  \begin{sideways}Exponent\end{sideways} &
  \begin{sideways}IsAbelian\end{sideways} &
  \begin{sideways}IsCyclic\end{sideways} &
  \begin{sideways}IsElementaryAbelian\end{sideways} &
  \begin{sideways}IsNilpotentGroup\end{sideways} &
  \begin{sideways}IsPerfectGroup\end{sideways} &
  \begin{sideways}IsPolycyclicGroup\end{sideways} &
  \begin{sideways}IsSupersolvableGroup$~$\end{sideways} &
  \begin{sideways}IsMonomialGroup\end{sideways} &
  \begin{sideways}IsSimpleGroup\end{sideways} &
  \begin{sideways}IsPGroup\end{sideways}
  \\
  \hline
  \hline
  $2$ & $1$ & $166$ & $166$ & $[ 2 ]$ & $2$  & Y & Y & Y & Y & N & Y & Y & Y & Y & Y\\
  \hline
  $3$ & $1$ & $31$ & $31$ & $[ 3 ]$ & $3$  & Y & Y & Y & Y & N & Y & Y & Y & Y & Y\\
  \hline
  $4$ & $1$ & $23$ & $23$ & $[ 4 ]$ & $4$  & Y & Y & N & Y & N & Y & Y & Y & N & Y\\
  $4$ & $2$ & $40$ & $40$ & $[ 2, 2 ]$ & $2$  & Y & N & Y & Y & N & Y & Y & Y & N & Y\\
  \hline
  $5$ & $1$ & $5$ & $5$ & $[ 5 ]$ & $5$  & Y & Y & Y & Y & N & Y & Y & Y & Y & Y\\
  \hline
  $6$ & $2$ & $4$ & $4$ & $[ 2, 3 ]$ & $6$  & Y & Y & N & Y & N & Y & Y & Y & N & N\\
  \hline
  $8$ & $1$ & $7$ & $7$ & $[ 8 ]$ & $8$  & Y & Y & N & Y & N & Y & Y & Y & N & Y\\
  $8$ & $2$ & $11$ & $11$ & $[ 2, 4 ]$ & $4$  & Y & N & N & Y & N & Y & Y & Y & N & Y\\
  $8$ & $4$ & $7$ & $7$ & $[ 2, 2 ]$ & $4$  & N & N & N & Y & N & Y & Y & Y & N & Y\\
  $8$ & $5$ & $1$ & $1$ & $[ 2, 2, 2 ]$ & $2$  & Y & N & Y & Y & N & Y & Y & Y & N & Y\\
  \hline
  $9$ & $2$ & $6$ & $6$ & $[ 3, 3 ]$ & $3$  & Y & N & Y & Y & N & Y & Y & Y & N & Y\\
  \hline
  $10$ & $2$ & $3$ & $3$ & $[ 2, 5 ]$ & $10$  & Y & Y & N & Y & N & Y & Y & Y & N & N\\
  \hline
  $12$ & $1$ & $2$ & $2$ & $[ 4 ]$ & $12$  & N & N & N & N & N & Y & Y & Y & N & N\\
  $12$ & $2$ & $2$ & $2$ & $[ 3, 4 ]$ & $12$  & Y & Y & N & Y & N & Y & Y & Y & N & N\\
  \hline
  $16$ & $2$ & $5$ & $3$ & $[ 4, 4 ]$ & $4$  & Y & N & N & Y & N & Y & Y & Y & N & Y\\
  $16$ & $4$ & $5$ & $5$ & $[ 2, 4 ]$ & $4$  & N & N & N & Y & N & Y & Y & Y & N & Y\\
  $16$ & $5$ & $5$ & $5$ & $[ 2, 8 ]$ & $8$  & Y & N & N & Y & N & Y & Y & Y & N & Y\\
  $16$ & $6$ & $2$ & $2$ & $[ 2, 4 ]$ & $8$  & N & N & N & Y & N & Y & Y & Y & N & Y\\
  $16$ & $10$ & $1$ & $1$ & $[ 2, 2, 4 ]$ & $4$  & Y & N & N & Y & N & Y & Y & Y & N & Y\\
  $16$ & $12$ & $5$ & $3$ & $[ 2, 2, 2 ]$ & $4$  & N & N & N & Y & N & Y & Y & Y & N & Y\\
  \hline
  $20$ & $5$ & $1$ & $1$ & $[ 2, 2, 5 ]$ & $10$  & Y & N & N & Y & N & Y & Y & Y & N & N\\
  \hline
  $25$ & $2$ & $2$ & $1$ & $[ 5, 5 ]$ & $5$  & Y & N & Y & Y & N & Y & Y & Y & N & Y\\
  \hline
  $32$ & $2$ & $1$ & $1$ & $[ 4, 4 ]$ & $4$  & N & N & N & Y & N & Y & Y & Y & N & Y\\
  $32$ & $3$ & $1$ & $1$ & $[ 4, 8 ]$ & $8$  & Y & N & N & Y & N & Y & Y & Y & N & Y\\
  $32$ & $4$ & $1$ & $1$ & $[ 4, 4 ]$ & $8$  & N & N & N & Y & N & Y & Y & Y & N & Y\\
  $32$ & $5$ & $1$ & $1$ & $[ 2, 8 ]$ & $8$  & N & N & N & Y & N & Y & Y & Y & N & Y\\
  $32$ & $13$ & $1$ & $1$ & $[ 2, 4 ]$ & $8$  & N & N & N & Y & N & Y & Y & Y & N & Y\\
  $32$ & $21$ & $1$ & $1$ & $[ 2, 4, 4 ]$ & $4$  & Y & N & N & Y & N & Y & Y & Y & N & Y\\
  $32$ & $23$ & $1$ & $1$ & $[ 2, 2, 4 ]$ & $4$  & N & N & N & Y & N & Y & Y & Y & N & Y\\
  $32$ & $35$ & $1$ & $1$ & $[ 2, 2, 2 ]$ & $4$  & N & N & N & Y & N & Y & Y & Y & N & Y\\
  $32$ & $47$ & $1$ & $1$ & $[ 2, 2, 2, 2 ]$ & $4$  & N & N & N & Y & N & Y & Y & Y & N & Y\\
  \hline
  $50$ & $5$ & $1$ & $0$ & $[ 2, 5, 5 ]$ & $10$  & Y & N & N & Y & N & Y & Y & Y & N & N\\
  \hline
  $64$ & $2$ & $1$ & $0$ & $[ 8, 8 ]$ & $8$  & Y & N & N & Y & N & Y & Y & Y & N & Y\\
  $64$ & $3$ & $1$ & $0$ & $[ 4, 8 ]$ & $8$  & N & N & N & Y & N & Y & Y & Y & N & Y\\
  $64$ & $68$ & $1$ & $0$ & $[ 2, 2, 4 ]$ & $4$  & N & N & N & Y & N & Y & Y & Y & N & Y\\
  $64$ & $72$ & $1$ & $0$ & $[ 2, 2, 4 ]$ & $4$  & N & N & N & Y & N & Y & Y & Y & N & Y\\
  $64$ & $179$ & $1$ & $0$ & $[ 2, 2, 2 ]$ & $8$  & N & N & N & Y & N & Y & Y & Y & N & Y\\
  \hline
\end{tabular}

  }
  \caption{Properties of the CICY groups.}
  \label{tab:groupprops}
\end{table}
An obvious question is whether we can guess any restrictions on
allowed groups by looking at the list of examples. General properties
of these groups are reviewed in \autoref{tab:groupprops}. Recall that,
for finite groups,
\begin{equation}
  \vcenter{\xymatrix@C=5mm@R=1mm{
      & 
      \text{polycyclic}
      \ar@{=>}[dl]
      \\
      \text{solvable} 
      &
      &
      \text{supersolvable}
      \ar@{=>}[ul]
      \ar@{=>}[dl]
      &
      \text{nilpotent}
      \ar@{=>}[l]
      &
      \text{Abelian}
      \ar@{=>}[l]
      &
      \text{cyclic}
      \ar@{=>}[l]
      .
      \\
      & 
      \text{monomial}
      \ar@{=>}[ul]
  }}
\end{equation}
Note that the dicyclic group quotient investigated
in~\cite{Braun:2009qy} is the only known non-nilpotent Calabi-Yau
group.
\begin{table}
  \centering
  \begin{tabular}{|c|cc|cc|c@{~~}c@{~~}c@{~~}c@{~~}c@{~~}c@{~~}c@{~~}c@{~~}c@{~~}c|}
  \hline
  Group & $|G|$ & ID & 
  \begin{sideways}Abelian invariants\end{sideways} &
  \begin{sideways}Exponent\end{sideways} &
  \begin{sideways}IsAbelian\end{sideways} &
  \begin{sideways}IsCyclic\end{sideways} &
  \begin{sideways}IsElementaryAbelian\end{sideways} &
  \begin{sideways}IsNilpotentGroup\end{sideways} &
  \begin{sideways}IsPerfectGroup\end{sideways} &
  \begin{sideways}IsPolycyclicGroup\end{sideways} &
  \begin{sideways}IsSupersolvableGroup$~$\end{sideways} &
  \begin{sideways}IsMonomialGroup\end{sideways} &
  \begin{sideways}IsSimpleGroup\end{sideways} &
  \begin{sideways}IsPGroup\end{sideways}
  \\
  \hline
  \hline
  $D_6$ & $6$ & $1$ & $[ 2 ]$ & $6$  & N & N & N & N & N & Y & Y & Y & N & N\\
  \hline
  $D_8$ & $8$ & $3$ & $[ 2, 2 ]$ & $4$  & N & N & N & Y & N & Y & Y & Y & N & Y\\
  \hline
  $\Z_9$ & $9$ & $1$ & $[ 9 ]$ & $9$  & Y & Y & N & Y & N & Y & Y & Y & N & Y\\
  \hline
  $D_{10}$ & $10$ & $1$ & $[ 2 ]$ & $10$  & N & N & N & N & N & Y & Y & Y & N & N\\
  \hline
  $A_4$ & $12$ & $3$ & $[ 3 ]$ & $6$  & N & N & N & N & N & Y & N & Y & N & N\\
  $\Z_6 \times \Z_2$ & $12$ & $5$ & $[ 2, 2, 3 ]$ & $6$  & Y & N & N & Y & N & Y & Y & Y & N & N\\
  \hline
  $\Z_{16}$ & $16$ & $1$ & $[ 16 ]$ & $16$  & Y & Y & N & Y & N & Y & Y & Y & N & Y\\
  $(\Z_4 \times \Z_2) \rtimes \Z_2$ & $16$ & $3$ & $[ 2, 4 ]$ & $4$  & N & N & N & Y & N & Y & Y & Y & N & Y\\
  $Q_{16}$ & $16$ & $9$ & $[ 2, 2 ]$ & $8$  & N & N & N & Y & N & Y & Y & Y & N & Y\\
  $\Z_2^4$ & $16$ & $14$ & $[ 2, 2, 2, 2 ]$ & $2$  & Y & N & Y & Y & N & Y & Y & Y & N & Y\\
  \hline
  $\Z_5 \rtimes \Z_4$ & $20$ & $1$ & $[ 4 ]$ & $20$  & N & N & N & N & N & Y & Y & Y & N & N\\
  $\Z_{20}$ & $20$ & $2$ & $[ 4, 5 ]$ & $20$  & Y & Y & N & Y & N & Y & Y & Y & N & N\\
  \hline
  $\Z_{25}$ & $25$ & $1$ & $[ 25 ]$ & $25$  & Y & Y & N & Y & N & Y & Y & Y & N & Y\\
  \hline
  $(\Z_2^2) . (\Z_4 \times \Z_2)$ & $32$ & $8$ & $[ 2, 4 ]$ & $8$  & N & N & N & Y & N & Y & Y & Y & N & Y\\
  $Q_8 \rtimes \Z_4$ & $32$ & $10$ & $[ 2, 4 ]$ & $8$  & N & N & N & Y & N & Y & Y & Y & N & Y\\
  $\Z_4 \rtimes \Z_8$ & $32$ & $12$ & $[ 2, 8 ]$ & $8$  & N & N & N & Y & N & Y & Y & Y & N & Y\\
  $\Z_8 \rtimes \Z_4$ & $32$ & $14$ & $[ 2, 4 ]$ & $8$  & N & N & N & Y & N & Y & Y & Y & N & Y\\
  $\Z_4 . D_8$ & $32$ & $15$ & $[ 2, 4 ]$ & $8$  & N & N & N & Y & N & Y & Y & Y & N & Y\\
  $\Z_4 \times Q_8$ & $32$ & $26$ & $[ 2, 2, 4 ]$ & $4$  & N & N & N & Y & N & Y & Y & Y & N & Y\\
  $(\Z_2^2) . (\Z_2^3)$ & $32$ & $32$ & $[ 2, 2, 2 ]$ & $4$  & N & N & N & Y & N & Y & Y & Y & N & Y\\
  $\Z_8 \times \Z_2 \times \Z_2$ & $32$ & $36$ & $[ 2, 2, 8 ]$ & $8$  & Y & N & N & Y & N & Y & Y & Y & N & Y\\
  $\Z_2 \times (\Z_8 \rtimes \Z_2)$ & $32$ & $37$ & $[ 2, 2, 4 ]$ & $8$  & N & N & N & Y & N & Y & Y & Y & N & Y\\
  \hline
  $\Z_{50}$ & $50$ & $2$ & $[ 2, 25 ]$ & $50$  & Y & Y & N & Y & N & Y & Y & Y & N & N\\
  \hline
\end{tabular}

  \caption{Groups that are \emph{not} in the list of freely acting
    CICY groups. That is, the freely acting groups on smooth CICYs can be
    characterized as the groups of order $2, 3, 4, 5, 6, 8, 9, 10, 12,
    16, 20, 25, 32, 50$ that do not contain one of the groups above as
    a subgroup.}
  \label{tab:nongroups}
\end{table}
In \autoref{tab:nongroups}, we describe the groups acting freely on
smooth CICYs by giving a list of subgroups that \emph{must} not
occur. As there is a limit of $|G|\leq 64$ just because of topological
indices, the forbidden subgroups of large order are presumably only an
artifact of the finite sample of Calabi-Yau threefolds under
consideration. 

However, it is a curious observation that the dihedral group $D_6$
with $6$ elements (a.k.a.\ the symmetric group on three letters $S_3$)
and the dihedral group $D_8$ are not allowed\footnote{In fact, only
  the exceptional dihedral groups $\Z_2=D_2$ and $\Z_2^2=D_4$ are
  allowed.}. Note that an ample divisor $D$ (that is, a divisor in the
dual of the K\"aher cone) in a Calabi-Yau threefold $X$ is a surface
of general type. By the Lefshetz hyperplane theorem
$\pi_1(D)=\pi_1(X)$. Focusing on the complete intersection of four
quadrics in $\CP^7$ (CICY $\#7861$), the minimal ample divisor is a
section of $\Osheaf(1)$, that is, a complete intersection of four
quadrics in $\CP^6$. Beauville~\cite{BeauvilleNonAbel} constructed a
free $Q_8$ action on this Calabi-Yau threefold and noted that the
$\Osheaf(1)$ divisor on the quotient is a so-called Campedelli
surface\footnote{A Campedelli surface $S$ is a surface of general type
  with $K^2=2$ and $h^0(S,K)=0$. Such a surface has $h^{11}(S)=8$ but
  its fundamental group is not uniquely determined. The size of the
  fundamental group is limited to $\pi_1(S)\leq 9$.} with
$\pi_1(D)=Q_8$. It is known~\cite{DanielNaie06011999, lopes-2008} that
Campedelli surfaces cannot have fundamental groups $D_{2n}$ for $n\geq
3$.

Of course Campedelli surfaces are the very exception amongst ample
divisors on CICYs. Moreover, any finite group can appear as the
fundamental groups of a surface of general type $S$ if one does not
pose any restriction on the Chern numbers $\int_S c_1^2$ and $\int_S
c_2$. Nevertheless, according to the classification result there are
no free $D_{2n}$-actions, $n\geq 3$, on any CICY.

\section*{Acknowledgments}

I would like to thank Philip Candelas, Rhys Davies, and Tony Pantev
for useful discussions. I also would like to thank Frank L\"ubeck,
Laurent Bartholdi, Willem de Graaf, and especially Alexander Hulpke
for their GAP support, and Hans Sch\"onemann for Singular support.

\appendix
% Hack that fixes reference name:
\makeatletter
\def\Hy@chapterstring{section}
\makeatother

\section{Group Cohomology}
\label{sec:groupcohomology}

The standard approach to a projective representation $r:G\to PGL(n)$
is by choosing a lift $\rt(g) \in GL(n)$ for each $g\in G$ and then
noting that there is a function
\begin{equation}
  c:G\times G\to \C^\times,
  \quad
  \rt(g) \rt(h) = c(g,h) \; \rt(gh) 
  \quad 
  \forall g,h\in G
  ,
\end{equation}
called the \emph{factor set}. Associativity implies that $c$ is a
$\C^\times$-valued cocycle, and multiplying the matrices $\rt(g)$ by
non-zero complex constants amounts to changing $c$ by a
coboundary. Therefore, the projective representation uniquely
determines a group cohomology\footnote{Group cohomology with
  coefficients $M$ is the usual (topological) cohomology of its
  classifying space, $H^2(G,M) \eqdef H^2(BG,M)$.} class $[c] \in
H^2(G,\C^\times)$.

A short exact sequence eq.~\eqref{eq:centralextension} defines a long
exact sequence in cohomology, 
\begin{equation}
  \cdots
  \longrightarrow
  H^1(\Gt,\C^\times)
  \stackrel{R}{\longrightarrow}
  H^1(K,\C^\times)
  \stackrel{\Delta}{\longrightarrow}
  H^2(G,\C^\times)
  \stackrel{S}{\longrightarrow}
  H^2(\Gt,\C^\times)
  \longrightarrow
  \cdots
\end{equation}
The maps $R$, $S$ are simply restriction (pull-back) via the maps in
the short exact sequence. Furthermore, note that $H^1(-,\C^\times) =
\Hom(-,\C^\times)$ are precisely the one-dimensional representations.

Now, a sufficient extension is one where $S=0$, that is, every factor
set of a projective $G$-representation pulls back to the trivial
factor set on $\Gt$. This is equivalent to $\Gt$ linearizing every
projective $G$-representation. In this case, 
\begin{equation}
  H^2(G,\C^\times) = \ker S = \img \Delta 
  = \coker R
  .
\end{equation}
But the cokernel of $R$ is precisely the set of twist classes in
Definition~\ref{def:twist}.

To summarize, the coboundary map $\Delta$ identifies twist classes
with the factor sets of projective representations as long as we have
chosen a sufficient extension $\Gt\to G$. If one chooses $\Gt$ too
small then the projective representations with $S\not=0$ cannot be
written as twisted representations.

\section{Character Formulas for SymInd/AltInd}
\label{sec:SymIndChar}

If one were to naively follow Definition~\ref{def:SymAltInd} in
evaluating the character-valued (anti-) symmetric induction
\begin{equation}
  \SymInd_H^G ,~ \AltInd_H^G
  :~
  \Hom(H,\C^\times) \longrightarrow \Hom(G,\C^\times)
\end{equation}
then one would have to first construct a representation for the given
$H$-character, compute the induced representation blocks, (anti-)
symmetrize, and then compute the trace to obtain the resulting
$G$-character. Obviously this is very inefficient, and we need an
equation that works on the level of group characters only.

The key to deriving such an equation is that, given a
$H$-representation $\gamma_1'$, the (anti-) symmetrized induction
$\SymInd_H^G(\gamma_1')$ is a sub-representation of $\Sym^{[G:H]}
\Ind_H^G(\gamma_1')$. Therefore, by subtracting the superfluous
representations, there must be a formula of the form
\begin{equation}
  \SymInd_H^G(\chi)
  = 
  \Sym^{[G:H]}\Big( \Ind_H^G(\chi)\Big)
  - \Big( \cdots \Big)
\end{equation}
only depending on the index $[G:H]$ of the subgroup $H$.  Using the
abbreviation $\Ind=\Ind_H^G$ and
\begin{equation}
  \Sym^{i_1,i_2,\dots,i_k}(\chi) = 
  \prod_j \Sym^{i_j}(\chi)
  ,\qquad
  \Alt^{i_1,i_2,\dots,i_k}(\chi) = 
  \prod_j \Alt^{i_j}(\chi)
  ,
\end{equation}
we find
\begin{subequations}
  \begin{align}
    \underline{[G:H]=1}: \qquad & \notag\\
    \SymInd(\chi) =&\;
    \Ind(\chi) = \chi
    ,\\
    \AltInd(\chi)
    =&\;
    \Ind(\chi) = \chi
    ,\displaybreak[0]\notag\\
    \underline{[G:H]=2}: \qquad & \notag\\
    \SymInd(\chi) =&\;
    \Sym^2 \Ind(\chi) - \Ind \Sym^2(\chi)
    ,\\
    \AltInd(\chi) =&\;
    \Alt^2\Ind(\chi) - \Ind  \Alt^2(\chi)
    ,\displaybreak[2]\notag\\ 
    \underline{[G:H]=3}: \qquad & \notag\\
    \SymInd(\chi) =&\;
    \Sym^3\Ind(\chi)
    - \Ind \Sym^3(\chi) 
    \notag\\ &\;
    - \Ind \Sym^2(\chi)    \Ind(\chi) 
    + \Ind \Sym^{2,1}(\chi)
    ,\\
    \AltInd(\chi) =&\;
    \Alt^3 \Ind(\chi) 
    - \Ind \Alt^3(\chi) 
    \notag\\ &\;
    - \Ind \Alt^2(\chi)     \Ind(\chi)
    + \Ind \Alt^{2,1}(\chi)
    ,\displaybreak[2]\notag\\
    \underline{[G:H]=4}: \qquad & \notag\\
    \SymInd(\chi) =&\;
    \Sym^4 \Ind(\chi)
    - \Sym^2 \Ind \Sym^2(\chi)
    - \Sym^2 \Ind(\chi)        \Ind \Sym^2(\chi)
    \notag\\ &\;
    + \Ind \Sym^{2,2}(\chi)
    - \Ind \Sym^{2,1,1}(\chi)
    - \Ind \Sym^3(\chi)        \Ind(\chi)
    \notag\\ &\;
    + \Ind \Sym^{2,1}(\chi)     \Ind(\chi)
    + \Ind \Sym^2(\chi)        \Ind \Sym^2(\chi)
    ,\\
    \AltInd(\chi) =&\;
    \Alt^4 \Ind(\chi)
    + \Alt^2 \Ind \Alt^2(\chi)
    - \Alt^2 \Ind(\chi)          \Ind \Alt^2(\chi)
    \notag\\ &\;
    + \Ind \Alt^{2,2}(\chi)
    - \Ind \Alt^{2,1,1}(\chi)
    - \Ind \Alt^3(\chi)          \Ind(\chi)
    \notag\\ &\;
    + \Ind \Alt^{2,1}(\chi)       \Ind(\chi)
    ,\displaybreak[2]\notag\\
    \underline{[G:H]=5}: \qquad & \notag\\
    \SymInd(\chi) =&\;
    \Sym^5 \Ind(\chi)
    - \Ind \Sym^5(\chi)
    + 2  \Sym^4 \Ind(\chi)         \Ind(\chi)
    \notag\\ &\;
    - \Sym^2 \Ind \Sym^2(\chi)   \Ind(\chi)
    - \Sym^3 \Ind(\chi)          \Sym^2 \Ind(\chi)
    \notag\\ &\;
    - \Sym^3 \Ind(\chi)          \Ind \Sym^2(\chi)
    - 9 \Ind \Sym^{4,1}(\chi)
    + \Ind \Sym^{3,2}(\chi)
    \notag\\ &\;
    + 19 \Ind \Sym^{3,1,1}(\chi)
    - \Ind \Sym^{2,2,1}(\chi)
    -12  \Ind\Sym^{3,1}(\chi)       \Ind(\chi)
    \notag\\ &\;
    - 9  \Ind\Sym^{2,1,1,1}(\chi)
    + \Ind \Sym^{2,1,1}(\chi)       \Ind(\chi)
    \notag\\ &\;
    + 2  \Ind\Sym^{1,1,1,1}(\chi)   \Ind(\chi)
    + \Ind\Sym^3(\chi)             \Ind\Sym^2(\chi)
    \notag\\ &\;
    + 6  \Ind\Sym^{2,2}(\chi)       \Ind(\chi)
    ,\\
    \AltInd(\chi) =&\;
    \Alt^5 \Ind(\chi)
    - \Ind(\Alt^5(\chi)
    + 2 \Alt^4 \Ind(\chi)          \Ind(\chi)
    \notag\\ &\;
    + \Alt^2 \Ind \Alt^2(\chi)   \Ind(\chi)
    - \Alt^3 \Ind(\chi)          \Alt^2 \Ind(\chi)
    \notag\\ &\;
    - \Alt^3 \Ind(\chi)          \Ind \Alt^2(\chi)
    - 9 \Ind \Alt^{4,1}(\chi)
    + \Ind \Alt^{3,2}(\chi)
    \notag\\ &\;
    + 19 \Ind \Alt^{3,1,1}(\chi)
    + 4 \Ind \Alt^{2,2,1}(\chi)
    -12 \Ind \Alt^{3,1}(\chi)       \Ind(\chi)
    \notag\\ &\;
    - 9 \Ind \Alt^{2,1,1,1}(\chi)
    + \Ind \Alt^{2,1,1} (\chi)       \Ind(\chi)
    \notag\\ &\;
    + 2 \Ind \Alt^{1,1,1,1} (\chi)   \Ind(\chi)
    + \Ind \Alt^3(\chi)             \Ind \Alt^2(\chi)
    .
    \notag
  \end{align}
\end{subequations}
Note that the formula for $\SymInd$ and $\AltInd$ are exactly
analogous for $[G:H]\leq 3$, but contain different coefficients for
$[G:H]\geq 4$.

\section{Guide to the Data Files}
\label{sec:data}

The complete list of free actions is available at
\url{http://www.stp.dias.ie/~vbraun/CICY/Quotients.tar.bz2}. Each
actions is contained in one of the 1695 files
\texttt{Data/ FreeQuotients/<CICY>-<Nr>.gap}, where \texttt{<CICY>} is
the CICY number, and \texttt{<Nr>} is an arbitrary and non-consecutive
labeling of different actions on the same CICY. The data files
themselves are GAP records with, hopefully, descriptive keywords and
can be read directly into GAP. As an example of how to use this
information, the GAP script \texttt{Data/LoadAction.gap} takes this
information and computes a basis for the invariant polynomials.

For example, let us look at the three-generation model studied
in~\cite{Braun:2009qy}:
\begin{Verbatim}[fontsize=\tiny,frame=single,rulecolor=\color{red}]
[vbraun@volker-desktop Data]$ gap.sh FreeQuotients/7246-21.gap LoadAction.gap
    
            #########           ######         ###########           ###  
         #############          ######         ############         ####  
        ##############         ########        #############       #####  
       ###############         ########        #####   ######      #####  
      ######         #         #########       #####    #####     ######  
     ######                   ##########       #####    #####    #######  
     #####                    ##### ####       #####   ######   ########  
     ####                    #####  #####      #############   ###  ####  
     #####     #######       ####    ####      ###########    ####  ####  
     #####     #######      #####    #####     ######        ####   ####  
     #####     #######      #####    #####     #####         #############
      #####      #####     ################    #####         #############
      ######     #####     ################    #####         #############
      ################    ##################   #####                ####  
       ###############    #####        #####   #####                ####  
         #############    #####        #####   #####                ####  
          #########      #####          #####  #####                ####  
                                                                          
     Information at:  http://www.gap-system.org
     Try '?help' for help. See also  '?copyright' and  '?authors'
    
   Loading the library. Please be patient, this may take a while.
GAP4, Version: 4.4.12 of 17-Dec-2008, x86_64-unknown-linux-gnu-gcc
Components:  small 2.1, small2 2.0, small3 2.0, small4 1.0, small5 1.0, small6 1.0, small7 1.0, small8 1.0, small9 1.0, small10 0.2, 
             id2 3.0, id3 2.1, id4 1.0, id5 1.0, id6 1.0, id9 1.0, id10 0.1, trans 1.0, prim 2.1  loaded.
Packages:    AutPGrp 1.4, GAPDoc 1.2, TomLib 1.1.4  loaded.
gap> FreeAction;
rec( CICY := rec( CICYmatrix := [ [ 0, 0, 1, 1, 1 ], [ 0, 0, 1, 1, 1 ], [ 1, 1, 0, 0, 1 ], [ 1, 1, 0, 0, 1 ] ], Pn := [ 3, 3, 3, 3 ],
      G := Group([ (1,2,3,5)(4,10,7,12)(6,11,9,8), (1,8,4)(2,10,6)(3,11,7)(5,12,9) ]), 
      GProw := [ (1,10,4,7)(2,11,6,9)(3,12,5,8), (1,3,2)(4,5,6)(7,9,8)(10,11,12) ] -> [ (1,3,2,4), () ], 
      GPcol := [ (1,10,4,7)(2,11,6,9)(3,12,5,8), (1,3,2)(4,5,6)(7,9,8)(10,11,12) ] -> [ (1,2), () ], Prow := Group([ (1,3,2,4), () ])
        , Pcol := Group([ (1,2), () ]), RowOrbit := [ [ 1, 2, 3, 4 ] ], RowOrbitFirst := [ 1 ], 
      RowOrbitFirstStabilizer := [ Group([ (1,2,3)(4,6,5)(7,8,9)(10,12,11) ]) ], ColOrbit := [ [ 1, 2 ], [ 3 ] ], 
      ColOrbitFirst := [ 1, 3 ], ColOrbitFirstStabilizer := [ Group([ (1,5,2,4,3,6)(7,11,8,10,9,12) ]), 
          Group([ (1,10,4,7)(2,11,6,9)(3,12,5,8), (1,3,2)(4,5,6)(7,9,8)(10,11,12) ]) ], DistinctEqns := [ [ 1, 2 ], [ 3, 4 ], [ 5 ] ]
        , Gcover := [ (1,10,4,7)(2,11,6,9)(3,12,5,8), (1,3,2)(4,5,6)(7,9,8)(10,11,12) ] -> [ (1,2,3,5)(4,10,7,12)(6,11,9,8), 
          (1,8,4)(2,10,6)(3,11,7)(5,12,9) ], K := Sym( [  ] ), TrivialTwists := [ Character( CharacterTable( Sym( [  ] ) ), [ 1 ] ) ]
        , Field := CF(12), Ring := CF(12)[x_1,x_2,x_3,x_4,x_5,x_6,x_7,x_8,x_9,x_10,x_11,x_12], 
      Coord := [ [ x_1, x_2, x_3, x_4, x_5, x_6, x_7, x_8, x_9, x_10, x_11, x_12 ] ], 
      CoordPn := [ [ x_1, x_2, x_3 ], [ x_4, x_5, x_6 ], [ x_7, x_8, x_9 ], [ x_10, x_11, x_12 ] ], 
      NontrivialConjClasses := [ (1,2,3)(4,6,5)(7,8,9)(10,12,11), (1,4)(2,6)(3,5)(7,10)(8,12)(9,11), (1,5,2,4,3,6)(7,11,8,10,9,12), 
          (1,7,4,10)(2,9,6,11)(3,8,5,12), (1,10,4,7)(2,11,6,9)(3,12,5,8) ], Num := 7246 ), 
  ChiGamma := [ Character( CharacterTable( Group([ (1,2,3)(4,6,5)(7,8,9)(10,12,11) ]) ), [ 3, 0, 0 ] ) ], 
  Gamma := [ [ (1,10,4,7)(2,11,6,9)(3,12,5,8), (1,3,2)(4,5,6)(7,9,8)(10,11,12) ] -> 
        [ [ [ 0, 0, 0, 0, 0, 0, 1, 0, 0, 0, 0, 0 ], [ 0, 0, 0, 0, 0, 0, 0, 1, 0, 0, 0, 0 ], [ 0, 0, 0, 0, 0, 0, 0, 0, 1, 0, 0, 0 ], 
              [ 0, 0, 0, 0, 0, 0, 0, 0, 0, 1, 0, 0 ], [ 0, 0, 0, 0, 0, 0, 0, 0, 0, 0, 1, 0 ], [ 0, 0, 0, 0, 0, 0, 0, 0, 0, 0, 0, 1 ],
              [ 0, 0, 0, 1, 0, 0, 0, 0, 0, 0, 0, 0 ], [ 0, 0, 0, 0, 1, 0, 0, 0, 0, 0, 0, 0 ], [ 0, 0, 0, 0, 0, 1, 0, 0, 0, 0, 0, 0 ],
              [ 1, 0, 0, 0, 0, 0, 0, 0, 0, 0, 0, 0 ], [ 0, 1, 0, 0, 0, 0, 0, 0, 0, 0, 0, 0 ], [ 0, 0, 1, 0, 0, 0, 0, 0, 0, 0, 0, 0 ] 
             ], [ [ 1, 0, 0, 0, 0, 0, 0, 0, 0, 0, 0, 0 ], [ 0, E(3)^2, 0, 0, 0, 0, 0, 0, 0, 0, 0, 0 ], 
              [ 0, 0, E(3), 0, 0, 0, 0, 0, 0, 0, 0, 0 ], [ 0, 0, 0, 1, 0, 0, 0, 0, 0, 0, 0, 0 ], 
              [ 0, 0, 0, 0, E(3)^2, 0, 0, 0, 0, 0, 0, 0 ], [ 0, 0, 0, 0, 0, E(3), 0, 0, 0, 0, 0, 0 ], 
              [ 0, 0, 0, 0, 0, 0, 1, 0, 0, 0, 0, 0 ], [ 0, 0, 0, 0, 0, 0, 0, E(3), 0, 0, 0, 0 ], 
              [ 0, 0, 0, 0, 0, 0, 0, 0, E(3)^2, 0, 0, 0 ], [ 0, 0, 0, 0, 0, 0, 0, 0, 0, 1, 0, 0 ], 
              [ 0, 0, 0, 0, 0, 0, 0, 0, 0, 0, E(3), 0 ], [ 0, 0, 0, 0, 0, 0, 0, 0, 0, 0, 0, E(3)^2 ] ] ] ], 
  ChiRho := [ Character( CharacterTable( Group([ (1,5,2,4,3,6)(7,11,8,10,9,12) ]) ), [ 2, -E(3), -E(3)^2, 2, -E(3)^2, -E(3) ] ), 
      Character( CharacterTable( Group([ (1,10,4,7)(2,11,6,9)(3,12,5,8), (1,3,2)(4,5,6)(7,9,8)(10,11,12) ]) ), [ 1, 1, 1, 1, 1, 1 
         ] ) ], 
  Rho := [ [ (1,10,4,7)(2,11,6,9)(3,12,5,8), (1,3,2)(4,5,6)(7,9,8)(10,11,12) ] -> [ [ [ 0, 0, 1, 0 ], [ 0, 0, 0, 1 ], [ 1, 0, 0, 0 ],
              [ 0, 1, 0, 0 ] ], [ [ 1, 0, 0, 0 ], [ 0, E(3), 0, 0 ], [ 0, 0, 1, 0 ], [ 0, 0, 0, E(3)^2 ] ] ], 
      [ (1,10,4,7)(2,11,6,9)(3,12,5,8), (1,3,2)(4,5,6)(7,9,8)(10,11,12) ] -> [ [ [ 1 ] ], [ [ 1 ] ] ] ], 
  Invariant := [ [ 0, 0, 0, 0, 0 ], [ 0, 0, 0, 0, x_3*x_6*x_9*x_12 ], 
      [ 0, 0, 0, 0, x_2*x_6*x_8*x_12+x_2*x_6*x_9*x_11+x_3*x_5*x_8*x_12+x_3*x_5*x_9*x_11 ], [ 0, 0, 0, 0, x_2*x_5*x_8*x_11 ], 
      [ 0, 0, 0, 0, x_1*x_6*x_8*x_11+x_2*x_5*x_7*x_12+x_2*x_5*x_9*x_10+x_3*x_4*x_8*x_11 ], 
      [ 0, 0, 0, 0, x_1*x_6*x_7*x_12+x_1*x_6*x_9*x_10+x_3*x_4*x_7*x_12+x_3*x_4*x_9*x_10 ], 
      [ 0, 0, 0, 0, x_1*x_5*x_9*x_12+x_2*x_4*x_9*x_12+x_3*x_6*x_7*x_11+x_3*x_6*x_8*x_10 ], 
      [ 0, 0, 0, 0, x_1*x_5*x_7*x_11+x_1*x_5*x_8*x_10+x_2*x_4*x_7*x_11+x_2*x_4*x_8*x_10 ], 
      [ 0, 0, 0, 0, x_1*x_4*x_8*x_12+x_1*x_4*x_9*x_11+x_2*x_6*x_7*x_10+x_3*x_5*x_7*x_10 ], [ 0, 0, 0, 0, x_1*x_4*x_7*x_10 ], 
      [ 0, x_9*x_12, 0, x_3*x_6, 0 ], [ 0, x_7*x_11+x_8*x_10, 0, x_1*x_5+x_2*x_4, 0 ], 
      [ x_8*x_12+x_9*x_11, 0, x_2*x_6+x_3*x_5, 0, 0 ], [ x_7*x_10, 0, x_1*x_4, 0, 0 ] ] )
\end{Verbatim}

The CICY configuration matrix is recorded as
\begin{equation}
  \begin{split}
    \vec{d} =&\;
    \text{\texttt{FreeAction.CICY.Pn}}
    ,\\
    c =&\;
    \text{\texttt{FreeAction.CICY.CICYmatrix}}
    ,\\
    \vec{\delta} =&\;
    \text{\texttt{List(FreeAction.CICY.DistinctEqns, Size)}}
  \end{split}
\end{equation}
and the CICY group, see Definition \ref{def:CICYgroup}, is
\begin{equation}
  \begin{split}
    C \;&= (d_i,c_{ij},\delta_j)_{i=1..n,~j=1..m}
    ,\\
    \Gt \;&= \text{\texttt{Source(FreeAction.CICY.Gcover)}}
    ,\\
    \pi_r \;&= \text{\texttt{FreeAction.CICY.GProw}}
    ,\\
    \pi_c \;&= \text{\texttt{FreeAction.CICY.GPcol}}
    .
  \end{split}
\end{equation}
Note that the group we are working with is always the (generalized)
Schur cover for the $\pi$-representation. The freely acting group on
the Calabi-Yau threefold is
\begin{equation}
  G 
  = \texttt{FreeAction.CICY.G}
  ,
  = \texttt{Image(FreeAction.CICY.Gcover)}
  .
\end{equation}
To entirely specify the CICY group action, we only need to specify two
(linear) $\pi$-representations of $\Gt$ acting on the homogeneous
coordinates and the polynomials. These are
\begin{equation}
  \begin{split}
    \gamma =&\;
    \text{\texttt{FreeAction.Gamma}}
    ,
    \\
    \rho =&\;
    \text{\texttt{FreeAction.Rho}}
    .
  \end{split}
\end{equation}
This is how the data file records the CICY group representation
$(C,G,\pi_r,\gamma,\pi_c,\rho)$, see
Definition~\ref{def:CICYgroupaction}. Finally, a set of
generators\footnote{Note that no attempt is made to return
  \emph{linearly independent} generators.} for the invariant
polynomials is stored in \texttt{FreeAction.Invariant}.

\bibliographystyle{utcaps} 
\renewcommand{\refname}{Bibliography}
\addcontentsline{toc}{section}{Bibliography} 
\bibliography{references}

\end{document}